\documentclass[iop]{emulateapj}

\usepackage{amsmath,amssymb,latexsym,graphics,epsfig,exscale,graphicx}
\usepackage{pstricks}
\usepackage{multirow}
\usepackage{graphicx}
\usepackage{subfigure}

\newcommand{\gad}{{\sc Gadget-2}}

\newcommand{\hmpc}{h^{-1}{\rm Mpc}}
\newcommand{\lcdm}{$\Lambda$CDM}
\newcommand{\Msun}{$M_{\odot}$}
\newcommand{\Msig}{$M_{\rm BH}$--$\sigma$}
\newcommand{\Mbulge}{$M_{\rm BH}$--$M_{\rm bulge}$}

\newcommand{\figsz}{0.36}
\newcommand{\figszc}{0.5}

\newcommand{\pathI}{figures}

\begin{document}

\title{Black Hole-Galaxy Correlations without Self-Regulation}
\shorttitle{Black Hole-Galaxy Correlations without Self-Regulation}

\shortauthors{D. Angl{\'e}s-Alc{\'a}zar, F. {\"O}zel, and R. Dav{\'e} }
\author{Daniel Angl{\'e}s-Alc{\'a}zar\altaffilmark{1},
              Feryal {\"O}zel\altaffilmark{2,3},
           Romeel Dav{\'e}\altaffilmark{2,4,5,6}}
\altaffiltext{1}{Department of Physics, University of Arizona, Tucson, 
AZ 85721, USA; {\rm anglesd@email.arizona.edu}}  
\altaffiltext{2}{Astronomy Department, University of Arizona, Tucson, 
AZ 85721, USA}        
\altaffiltext{3}{Radcliffe Institute for Advanced Study, Harvard University, 
Cambridge, MA 02138, USA}
\altaffiltext{4}{University of the Western Cape, Bellville, Cape Town 7535, South Africa}
\altaffiltext{5}{South African Astronomical Observatories, Observatory, Cape Town 7925, South Africa}
\altaffiltext{6}{African Institute for Mathematical Sciences, Muizenberg, Cape Town 7945, South Africa}

\begin{abstract}
 
Recent models of black hole
growth in a cosmological context have forwarded a paradigm in which
the growth is self-regulated by feedback from the black hole itself.
Here we use cosmological zoom simulations of galaxy
formation down to $z = 2$ to show that such strong self-regulation is
required in the popular spherical Bondi accretion model, 
but that a plausible alternative model in which black
hole growth is limited by galaxy-scale torques does not
require self-regulation.  Instead, this torque-limited accretion model
yields black holes and galaxies evolving on average along the observed
scaling relations by relying only on a fixed, 5\,\% mass retention
rate onto the black hole from the radius at which the accretion flow
is fed.  Feedback from the black hole may (and likely does) occur, but
does not need to couple to galaxy-scale gas in order to regulate black
hole growth.  We show that this result is insensitive to variations in
the initial black hole mass, stellar feedback, or other implementation
details.  The torque-limited model allows for high accretion rates at
very early epochs (unlike the Bondi case), which if viable can help
explain the rapid early growth of black holes, while by $z \sim 2$ it
yields Eddington factors of $\sim 1$\%--10\%.  This model also yields
a less direct correspondence between major merger events and rapid
phases of black hole growth.  Instead, growth is more closely tied to
cosmological disk feeding, which may help explain observational
studies showing that, at least at $z \gtrsim 1$, active galaxies do
not preferentially show merger signatures.

\end{abstract}



\section{Introduction}

There is by now ample evidence that supermassive black holes reside at
the center of most sizeable galaxies \citep[e.g.,][]{ferr05}.  During
the last two decades, observations of local galaxies have established
a number of well-defined relationships between the mass of the central
black hole and stellar properties of the host galaxy such as the
mass of the central bulge \citep[\Mbulge~relation;][]{mago98,har04},
the concentration or S{\'e}rsic index \citep{gra01}, the bulge
velocity dispersion
\citep[\Msig~relation;][]{ferr00,trem02,gulte09,macc11}, the virial
bulge mass \citep{mar03}, the spheroid binding energy \citep{alle07},
and even the mass of the dark matter halo \citep{ferr02}.  The
tightness of these correlations suggests that there is a close
connection between the formation and evolution of galaxies and their
central black holes.  Moreover, the similarity between the global cosmic star
formation history \citep{mad96} and the evolution of quasar abundances
\citep{boy98}, as well as the apparent connection of starburst events
and active galactic nuclei (AGNs) in individual objects \citep{kau03}
provide further support for a link between galaxy formation and
accretion onto a central black hole.  Understanding this connection can have
broad implications in galaxy evolution and the formation of massive
black holes in the early universe.

Analytic arguments suggest that the observed black hole--galaxy correlations
might arise from the coupling of energy and/or momentum from radiation
emitted by the black hole accretion process to the surrounding galaxy (a
process known as black hole or AGN feedback), leading to the idea of
self-regulated growth and the co-evolution of black holes and galaxies
\citep{sil98,fab99,kin03,beg05,mur05,mcq12}.  The energy produced by
accretion onto the black hole can easily exceed the binding energy of its
parent galaxy and, therefore, even if a small fraction of this energy
is transferred to the surrounding gas, black hole feedback can have a profound
effect on the evolution of its host galaxy \citep{fab12}.  This
interpretation has been particularly attractive for galaxy formation
models since both semianalytic models and cosmological hydrodynamic
simulations require the injection of additional energy to prevent the
overcooling of gas and the subsequent runaway formation of stars
\citep[e.g.,][]{cro06,som08,gab11,tey11}.  AGN feedback is often
invoked as the source of this energy.

Motivated by these ideas, an increasing number of studies have now
incorporated models for black hole growth and feedback into idealized galaxy
scale simulations as well as full cosmological simulations
\citep[e.g.,][]{spr05b,dimat08,boo09,deb11,kim11,pow11,cho12,dub12}.
Since the numerical resolution required to self-consistently follow
gas inflows from $\sim 10$\,kpc host-galaxy scales down to the central
black hole ($\sim 10^{-6}$\,pc for the Schwarzschild radius of a
$10^{7}$\,\Msun~black hole) is out of reach for current computers, simulations
must rely on ``sub-grid" models to infer black hole accretion rates based on
the physical properties of the surrounding gas at the scales resolved
in the simulation.  The majority of galaxy formation simulations to
date have employed black hole accretion prescriptions based on the spherical
Bondi--Hoyle--Littleton parameterization \citep[hereafter ``Bondi
rate";][]{hoy39,bon44,bon52}, with the exception of some recent models
that incorporate the physics of angular momentum transport in
non-spherically symmetric flows \citep[e.g.,][]{deb11,pow11}.  A
variety of feedback models have been implemented to date, including
the injection of a fraction of the accretion luminosity as thermal
energy into the gas surrounding the black hole \citep{spr05b,sij07,joh09}, the
injection of momentum from optically thick radiation fields
\citep{deb11}, heating and radiation pressure from X-ray generated by
the AGN \citep{ham11,kim11,cho12}, and the injection of mass and/or
momentum from AGN-driven winds or radio jets
\citep{cho12,deb12,dub12}.

The success of coupled black hole accretion and feedback models in accounting
for a number of observations including black hole mass functions, Eddington
ratio distributions, and black hole--galaxy correlations
\citep[e.g.,][]{dimat05,hop06} have contributed to establish a
paradigm in which black hole growth is self-regulated by feedback from the black hole
itself.  There is, however, no direct observational evidence to date
for such self-regulated growth of black holes due to their own feedback
\citep{alex12}.  AGN feedback acting in the kinetic or radio mode is
commonly observed in the form of radio-emitting relativistic jets, and
its effects on the hot gas in galaxies and clusters are relatively
well understood.  The radiative or quasar feedback mode has been
identified by the presence of highly blueshifted absorption and
emission lines and broad line wings, but the overall impact of
quasar-driven winds on the host galaxy remains uncertain
\citep{fab12}.  Even if AGN feedback is effective in quenching star
formation in galaxies and suppressing cooling flows in groups and
clusters (as required by galaxy formation models), there is a priori
no reason for AGN feedback to be the dominant physical process
regulating black hole growth in galaxies.

In the presence of a non-zero net angular momentum of the surrounding
gas, black hole feeding likely occurs through viscous transport in a Keplerian
accretion disk \citep[e.g.,][]{shak73,bal98}.  Therefore, it is
plausible that most of the feedback energy and momentum escapes along
the polar direction without significantly affecting further accretion
\citep[e.g.,][]{ohs05}.  In order to allow for significant black hole growth,
the accretion disk must be continuously replenished by gas inflows from
larger scales.  The angular momentum of the gas along with competition
with star formation are known to represent significant barriers to the
transport of gas from galactic scales down to the black hole accretion disk
\citep[e.g.,][]{tho05,jog06,hop10}.  Thus, the rate of angular
momentum transport relative to the local star formation rate (SFR) may
well be the dominant physical process governing black hole growth in galaxies,
even if AGN feedback is acting at some level
\citep[e.g.,][]{esc06,esc07,pow11}.

Numerical simulations have shown that large-scale tidal torques
produced by galaxy interactions, galaxy mergers, and gravitational
instabilities in self-gravitating disks can lead to efficient angular
momentum transport and the rapid inflow of gas into the central
kiloparsec of galaxies \citep[e.g.,][]{her89,shl89,bar92}.  At smaller
scales, large-scale torques become less efficient and additional
mechanisms are required.  Using multiple nested galaxy scale
simulations of progressively higher resolution, \citet{hop10} showed
that a series of gravitational instabilities can generate net inflow
rates of up to 1--10\,\Msun\,yr$^{-1}$ down to sub-parsec scales
($\lesssim 0.1$\,pc).  They found that non-axisymmetric perturbations
to the stellar gravitational potential produce orbit crossings and
shocks in the gas that can be efficient in removing angular momentum
even at scales $\lesssim 10$\,pc.  \citet{hop11} derived analytic
expressions for the loss of angular momentum in the presence of such
shocks and presented an analytical ``gravitational torque" model for
the resulting gas inflow rates.  This analytic model was found to
reproduce gas inflow rates at $\lesssim 0.1$\,pc scales from
simulations with significantly less scatter than the Bondi
parameterization for the total enclosed mass within a given radius.

The goal of this paper is to investigate the growth of supermassive
black holes due to gravitational-torque-driven accretion in a cosmological
context and determine whether this growth mechanism can account for
the observed black hole--galaxy correlations without the requirements of AGN
feedback and self-regulated growth.  Our approach consists on
combining high-resolution cosmological hydrodynamic simulations
together with analytic models of black hole accretion in post-processing.
This simplification allows us to follow the growth of galaxies and black holes
from early epochs down to $z = 2$ without making any prior assumptions
about the effects of AGN feedback on galactic scales.  By comparing
predictions from the Bondi and gravitational torque models, we evaluate
the relative importance of feedback and angular momentum transport in
regulating cosmological black hole growth.  In addition, we discuss the
implications of different black hole accretion models on the link between
major galaxy mergers and rapid phases of black hole growth, as well as the
growth of black hole seeds in the early universe.  The observed black hole--galaxy
correlations are regarded here as a strong constraint for theories of
black hole growth and galaxy evolution but not as a mandatory outcome of AGN
feedback, as it is commonly assumed.  Given that both star formation
and black hole accretion histories peak at $z \sim 2$, our simulations target
the critical epoch when such correlations were likely established.

This paper is organized as follows.  We describe our simulations and
main analysis procedures in Section~\ref{sec:sim}.  In
Section~\ref{sec:galpro}, we present an overview of our simulated
galaxies and describe their main physical properties.  In
Section~\ref{sec:acc}, we evaluate the Bondi and gravitational torque
models and show how the inferred black hole accretion rates relate to the
physical properties and evolution of the host galaxies.  In
Section~\ref{sec:growth}, we use the observed black hole--galaxy correlations
to put constraints on black hole accretion in cosmological timescales and
evaluate the effects of AGN feedback as required by different black hole
models.  In Sections~\ref{sec:newmod} and~\ref{sec:corr}, we show that
the gravitational torque model naturally yields the observed
black hole--galaxy correlations regardless of the initial masses of black holes and
galaxies, and without the need for self-regulation by galaxy-scale AGN
feedback.  Finally, we summarize our results in Section~\ref{sec:end}.


\section{Simulations}\label{sec:sim}

We use the extended version of the \gad~code \citep{spr05} that has
been fully described in \citet{opp08}.  \gad~combines a TreePM
algorithm for computing gravitational forces with an
entropy-conserving formulation of smoothed particle hydrodynamics
\citep[SPH;][]{spr02}.  We include radiative cooling from primordial gas
assuming ionization equilibrium as in \citet{kat96} and metal-line
cooling based on \citet{sut93}.  Photoionization heating is modeled
via a spatially uniform, optically thin UV background \citep{haa01}
starting at $z = 9$.

Star formation and feedback are modeled in \gad~through a sub-grid
prescription.  Gas particles that are sufficiently dense to become
Jeans unstable are treated as a multi-phase fluid with cold, dense
clumps embedded within a hot phase medium \citep{mck77,spr03}.  Stars
form from the cold phase and feedback from Type II supernovae causes
evaporation of cold clumps into the hot phase (which can condense back
onto cold clumps) and metal enrichment.  The resulting SFRs are in
accord with the observed \citet{ken98} relation.  We also account for
energy and metal feedback from Type Ia supernovae and mass loss and
metal enrichment from asymptotic giant branch stars, tracking the
production of four metal species (C, O, Si, and Fe) separately as
described in \citet{opp08}.  In addition, our simulations include a
galactic outflow mechanism that imparts kinetic energy to gas
particles.  The outflow rates scale with galactic velocity dispersion,
$\sigma$, and the ratio of material entering the wind versus forming
stars (i.e., the mass loading factor) scales as $1/\sigma$, as in the
momentum-driven wind case \citep{mur05}.  An on-the-fly galaxy finder
is used to compute galaxy masses which are converted to $\sigma$ using
standard relations \citep{mo98}.

Overall, this treatment of metal production and galactic winds has
been successful in reproducing a broad range of observations including
the chemical enrichment of the intergalactic medium at $z > 2$ \citep{opp06,opp08}, the
luminosity function of high-redshift galaxies \citep{dav06,fin07}, and
the galaxy mass--metallicity relation \citep{fin08,dav11}.  In
addition, the high-resolution zoom simulations presented here have
been successful in reproducing many of the star formation,
morphological, and kinematic properties of $z = 2$ galaxies
\citep{ang12}.

We note that our simulations do not include prescriptions for black hole
growth and the effects of AGN feedback. Our goal is to use galaxy
formation simulations together with analytic black hole accretion models to
study the physical processes governing cosmological black hole growth, based
only on the observed black hole--galaxy correlations, and with no implicit
assumptions about the amount and efficiency of feedback from the
accretion process.

\subsection{Simulation Runs}

In this work we use the ``zoom-in" technique \cite[e.g.,][]{nav94} to
run cosmological simulations and resolve the physical conditions at
the inner kiloparsec of galaxies over cosmic time.  We first ran an
intermediate-resolution cosmological simulation with $2 \times
256^{3}$ gas+dark matter particles in a $[24\,\hmpc]^3$ comoving box.
This simulation allows us to characterize the population from which
galaxies are selected for resimulation.  We chose two galaxies at $z =
2$ characterized by similar masses but different morphologies,
environments, and merger histories.  We identified all particles
within the virial radius of each galaxy and traced them back to their
locations on the initial grid, where they mark the refinement region.
Zoom initial conditions were then generated by populating the
refinement region with a larger number of lower mass particles (each
particle is replaced by $8^{l}$, where $l$ is the zoom level), and
adding the additional small-scale fluctuations appropriate to the new
Nyquist frequency.  In order to reduce numerical artifacts due to the
difference in particle masses and to make sure that the large-scale
gravitational torques acting on the target halos are accurately
represented, the high-resolution region was significantly enlarged by
an iterative ``cleaning" procedure and surrounded by two nested,
concentric layers of progressively lower resolution.  After running
the simulations, we enlarged our galaxy sample by including six
additional galaxies that were found well within the high-resolution
regions, confirming that there is no contamination of low-resolution
particles within the virial radius of each galaxy.

Our zoom simulations have an effective resolution equivalent to $2
\times 1024^3$ particles homogeneously distributed in a
$[24\,\hmpc]^3$ box, with (high-resolution) gas particle mass $m_{\rm
gas} \approx 2.3 \times 10^5$\,\Msun, dark matter particle mass
$m_{\rm DM} \approx 1.2 \times 10^6$\,\Msun~and softening length
$\epsilon \approx 0.47\,h^{-1}$\,kpc comoving (or $\sim 224$\,pc
physical at $z = 2$).  All results presented in this paper (except
where indicated) correspond to simulations including our preferred
model for galactic outflows, based on momentum-conserving scalings as
explained above.  In order to identify the possible effects of winds
in our results, we also ran all simulations without winds.
Additionally, we ran simulations with $2 \times 512^3$ effective
resolution (i.e., two times lower spatial resolution and eight times lower
mass resolution) for all galaxies to test for numerical convergence.

Throughout this paper we assume a \lcdm~cosmology with parameters
$\Omega_{\rm \Lambda} = 0.72$, $\Omega_{\rm M} = 0.28$, $\Omega_{\rm
b} = 0.046$, $h = 0.7$, $\sigma_{8} = 0.82$, and $n = 0.96$,
consistent with the five-year {\it Wilkinson Microwave Anisotropy Probe} concordance cosmology
\citep{kom09}.

\begin{figure*}
\begin{center}
\includegraphics[scale=\figsz]{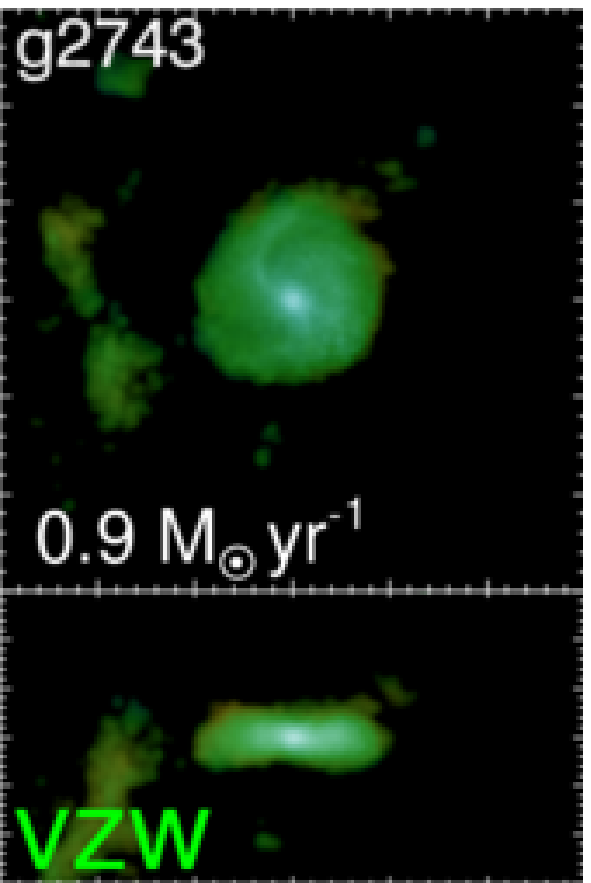}
\includegraphics[scale=\figsz]{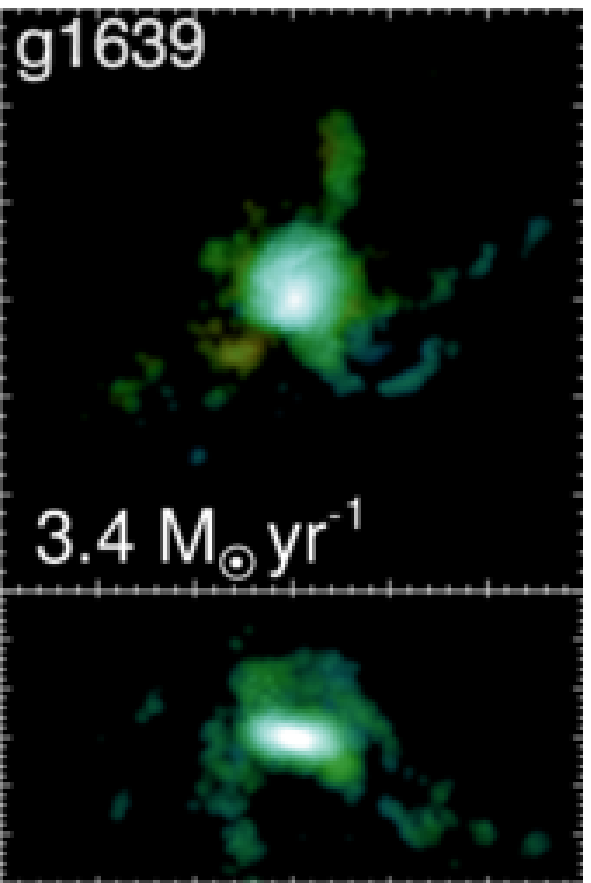}
\includegraphics[scale=\figsz]{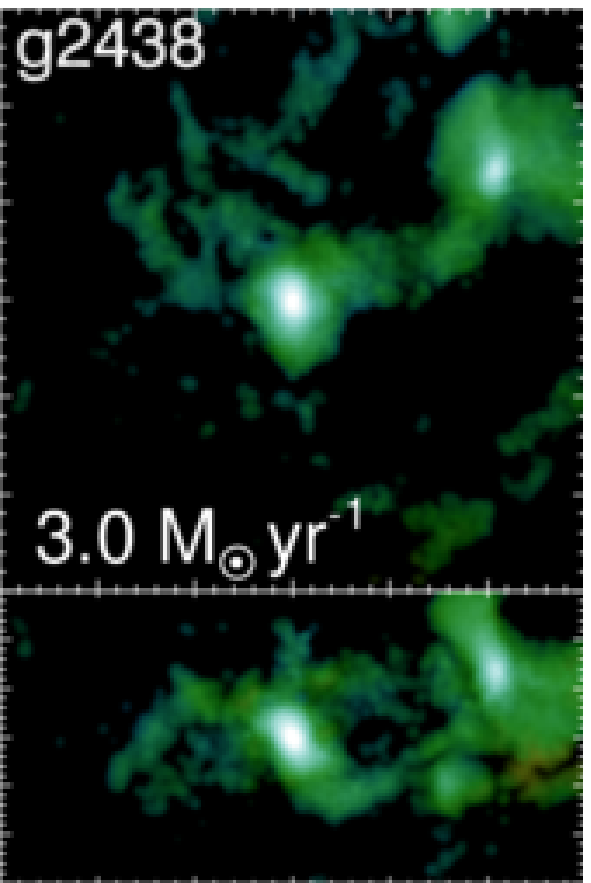}
\includegraphics[scale=\figsz]{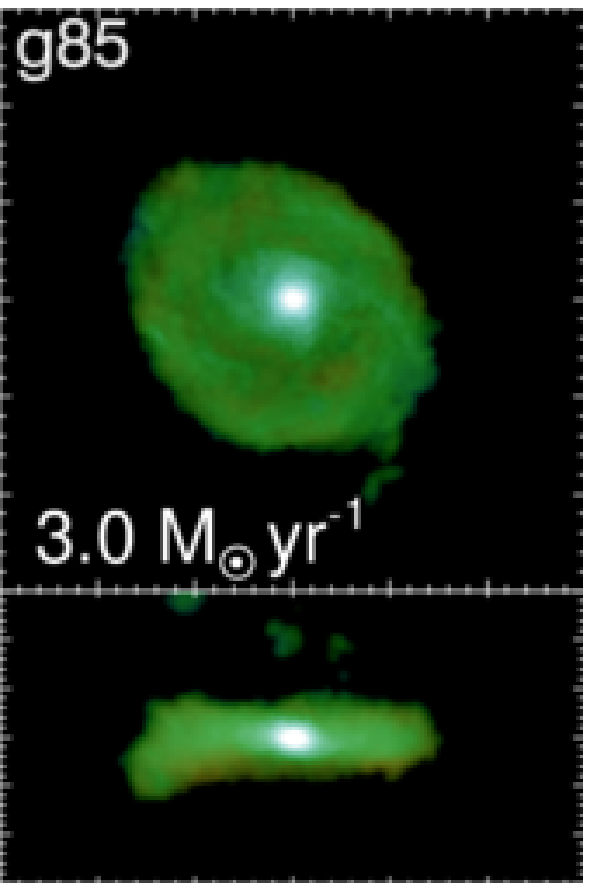}
\includegraphics[scale=\figsz]{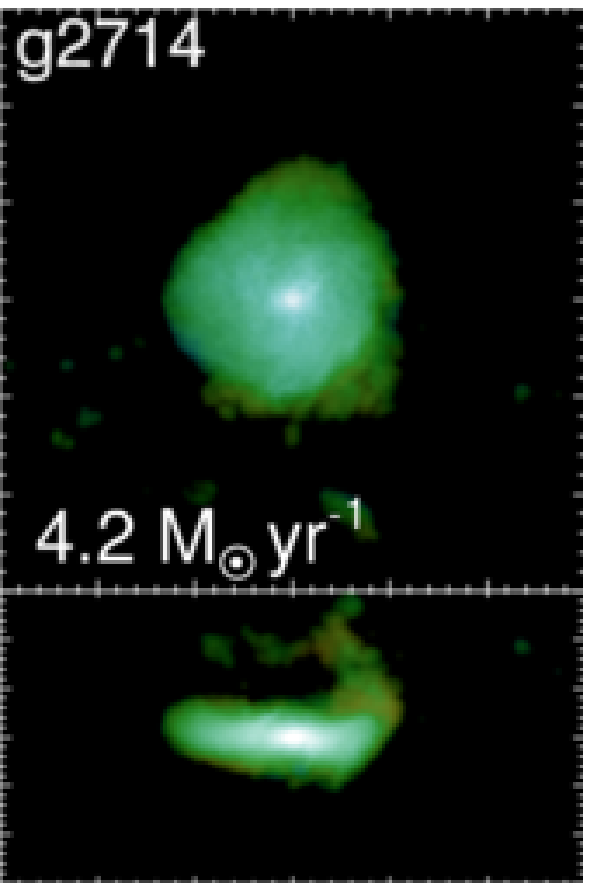}
\includegraphics[scale=\figsz]{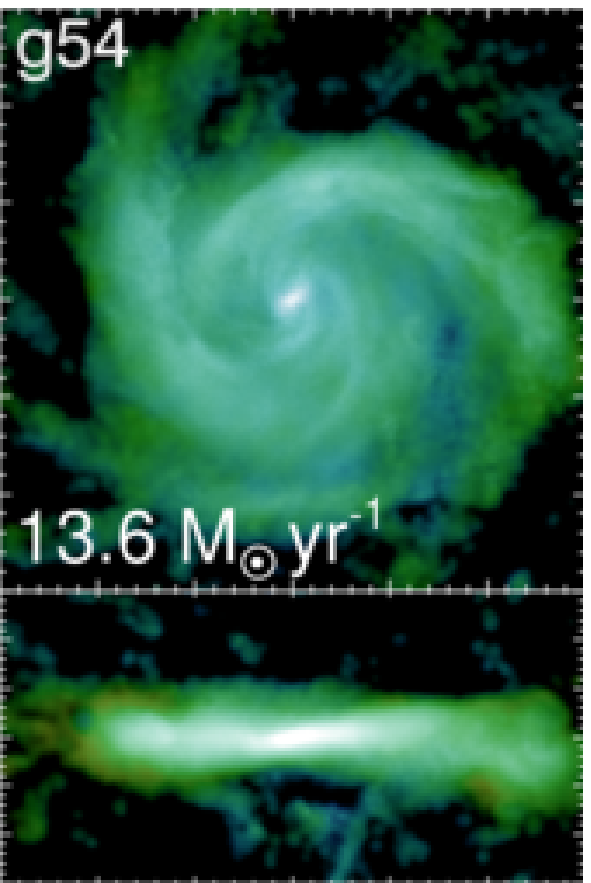}
\includegraphics[scale=\figsz]{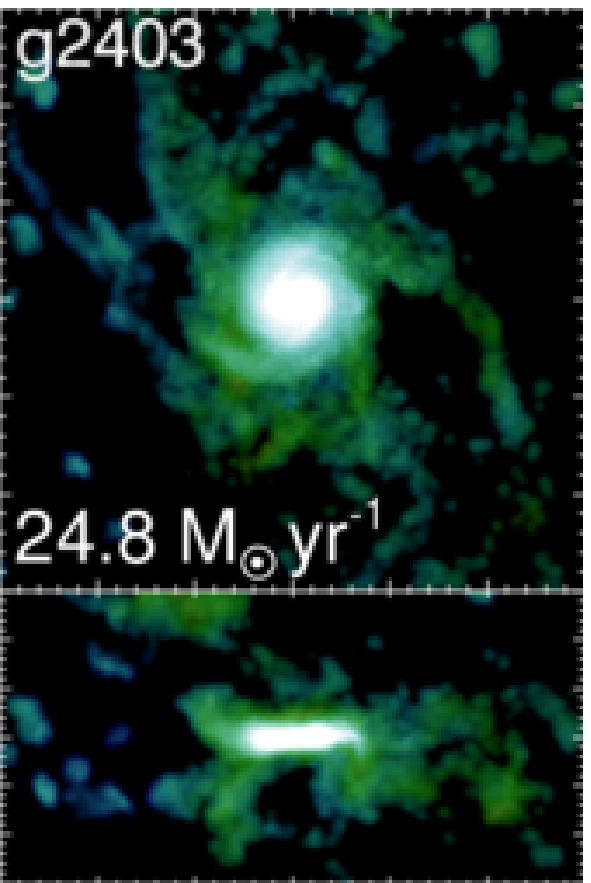}
\includegraphics[scale=\figsz]{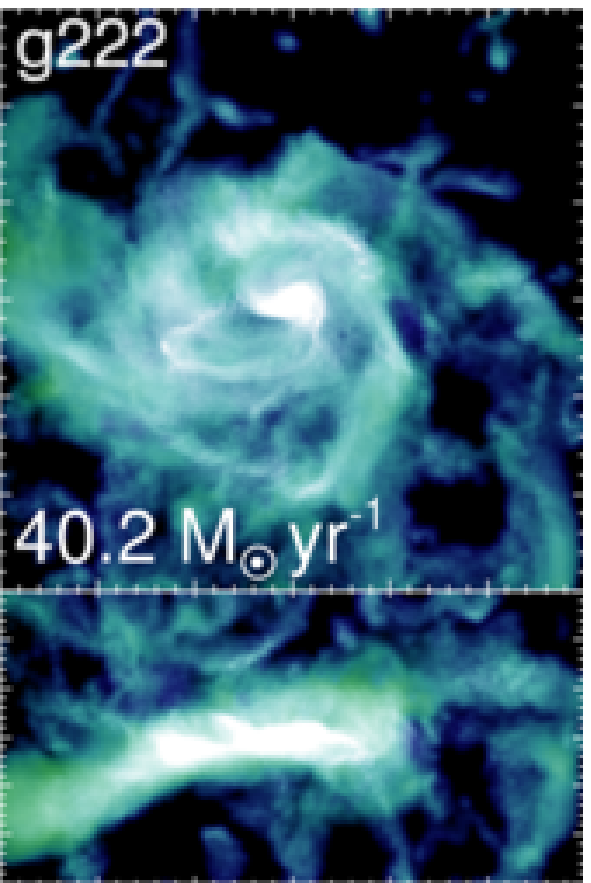}

\includegraphics[scale=\figszc]{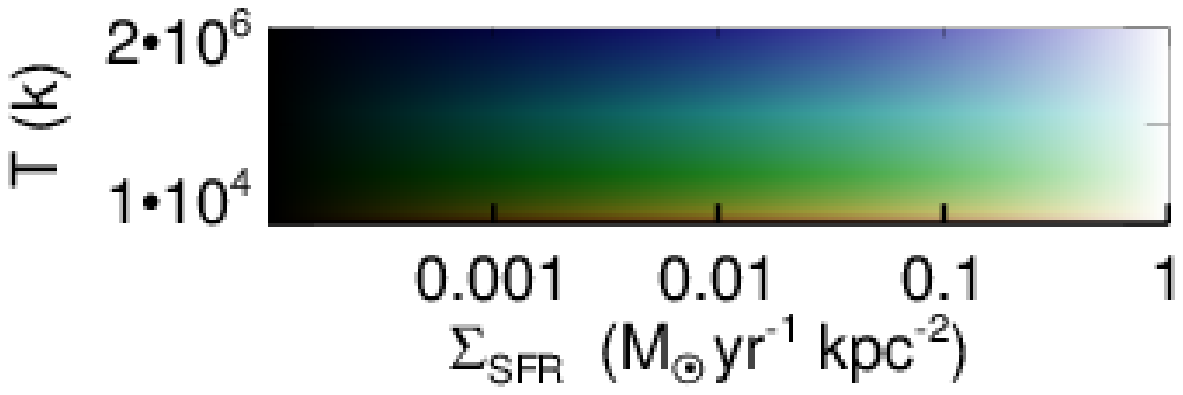}
\includegraphics[scale=\figszc]{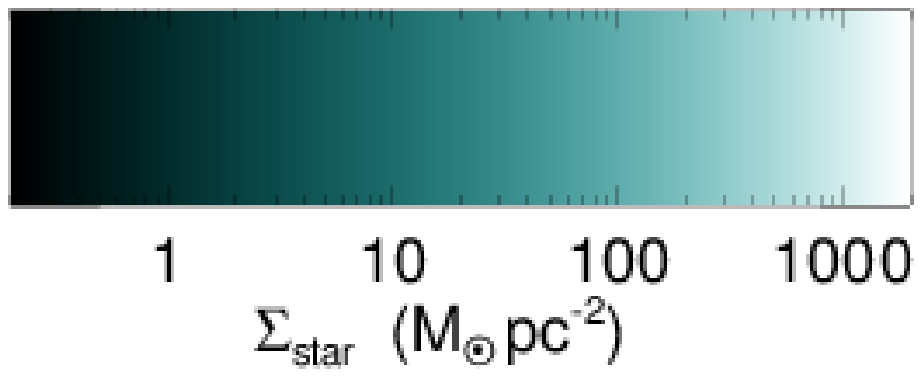}

\includegraphics[scale=\figsz]{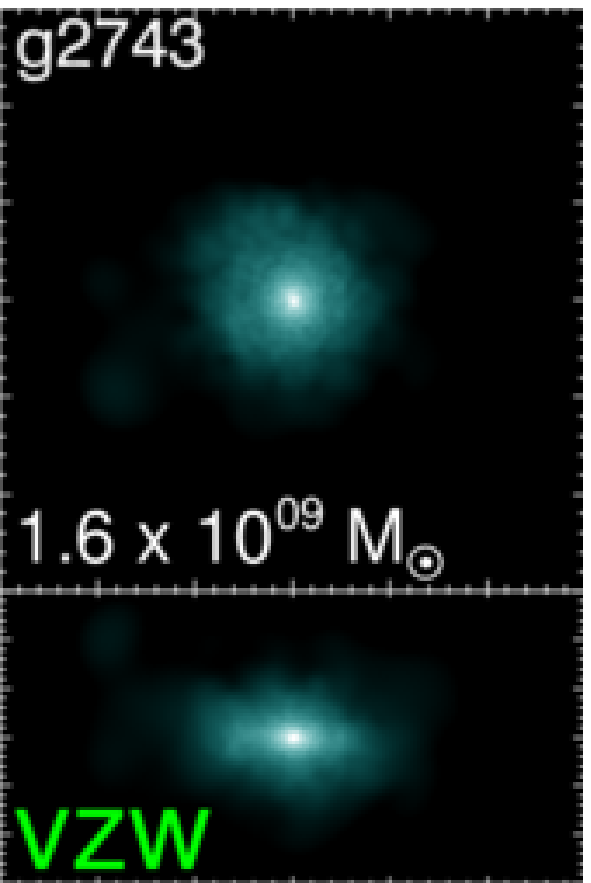}
\includegraphics[scale=\figsz]{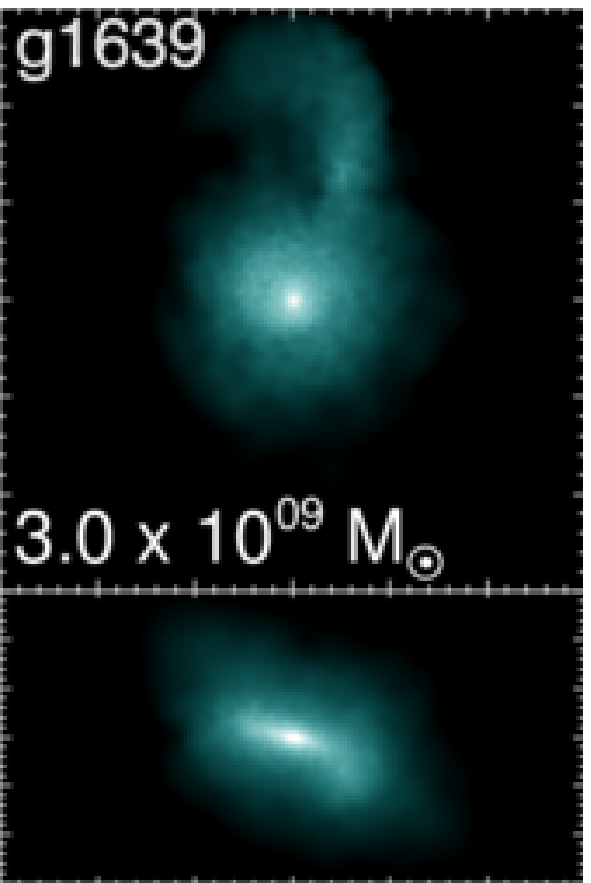}
\includegraphics[scale=\figsz]{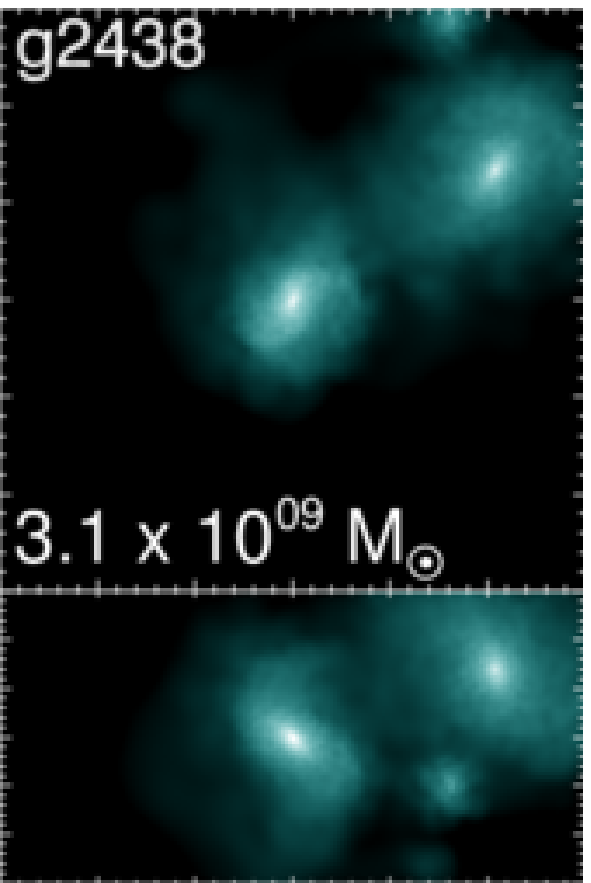}
\includegraphics[scale=\figsz]{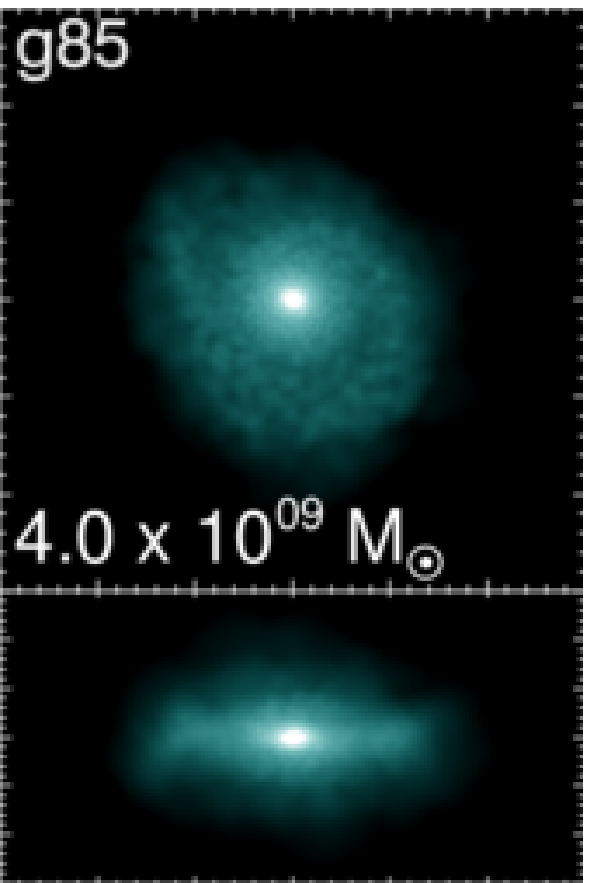}
\includegraphics[scale=\figsz]{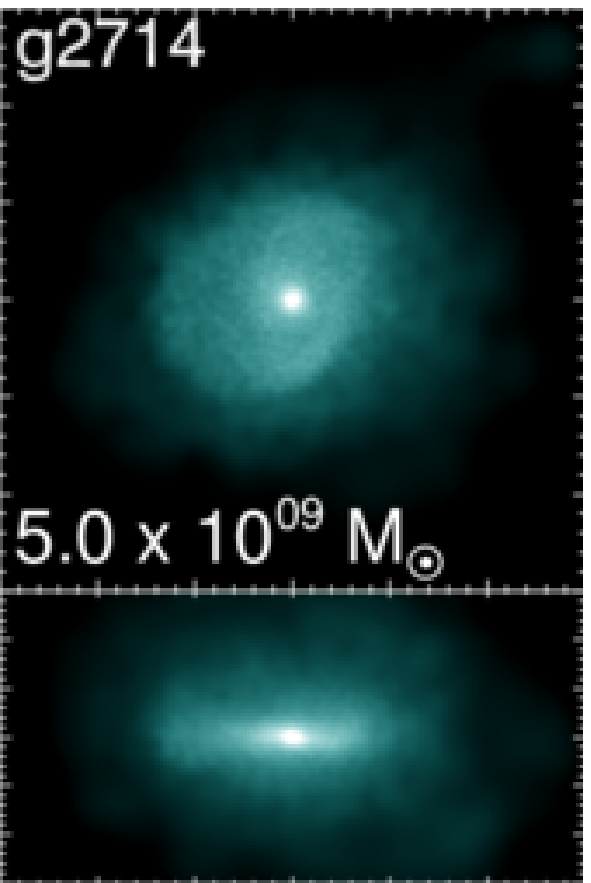}
\includegraphics[scale=\figsz]{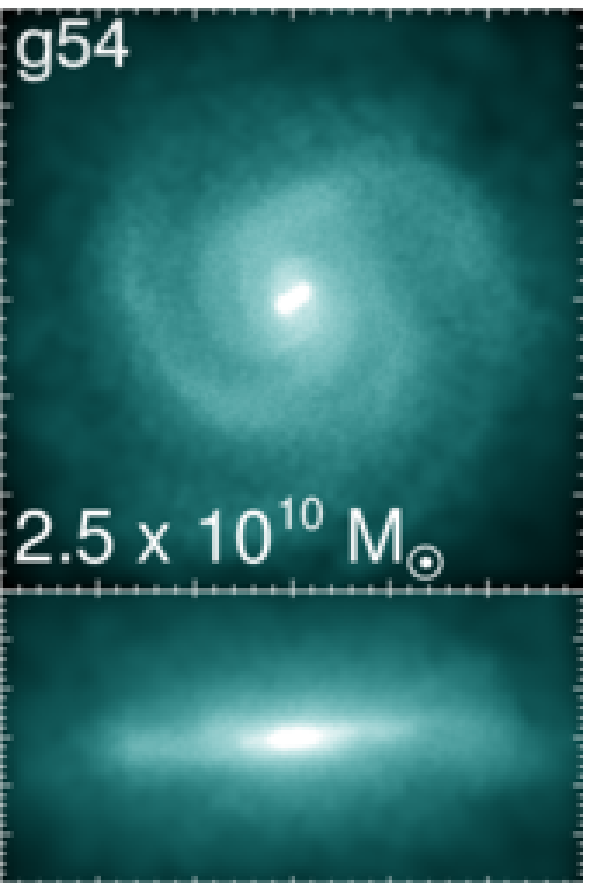}
\includegraphics[scale=\figsz]{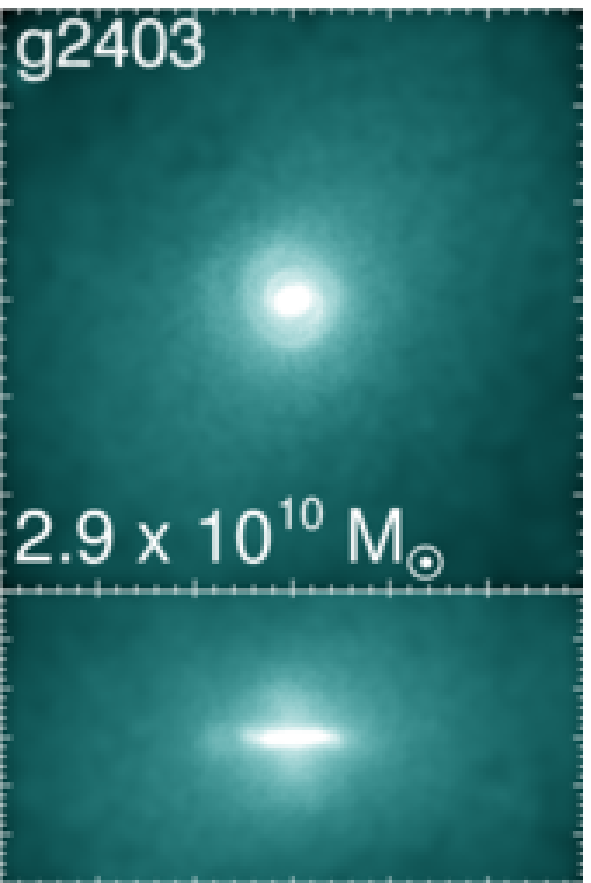}
\includegraphics[scale=\figsz]{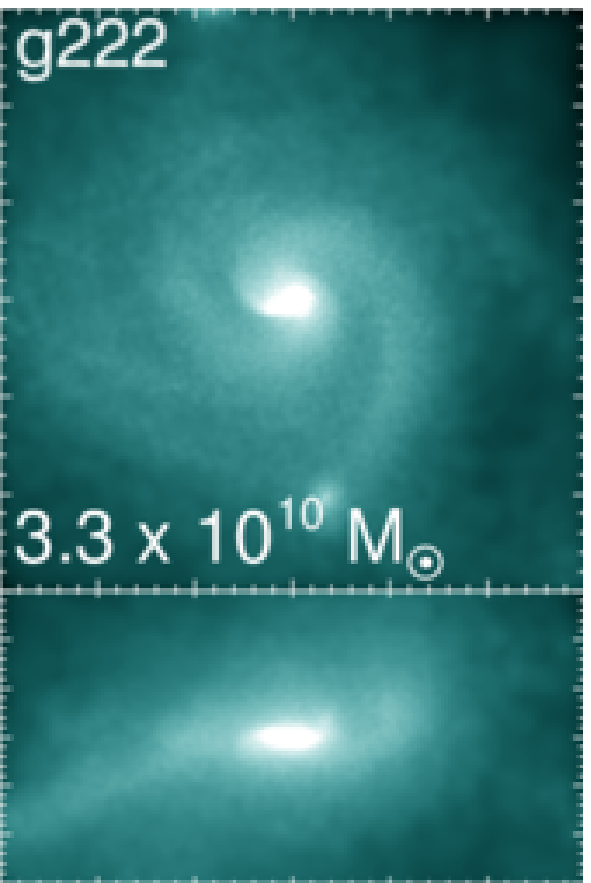}
\end{center}
\caption{Snapshots of our sample of simulated galaxies at $z = 2$.  
For each galaxy, we show SFR surface density maps
color-coded according to the projected density-weighted temperature
(top panels) and the projected stellar surface density (bottom
panels).  The total SFRs and stellar masses of
galaxies are indicated in the top and bottom panels, respectively.
Face-on and edge-on views of galaxies correspond to the direction of
the angular momentum at $z = 2$. The region shown is 30\,kpc
(physical) across.}
\label{fig:snaps}
\end{figure*}

\subsection{Analysis of Simulations}

We produced over 230 snapshot files at time intervals ranging from
$\sim 5$ to 25\,Myr during the simulation.  We identify galaxies by
means of the Spline Kernel Interpolative Denmax algorithm ({\sc
skid}\footnote[1]{http://www-hpcc.astro.washington.edu/tools/skid.html})
at each redshift snapshot independently \citep[see][for a
description]{ker05}.  Our simulated galaxies are thus defined as bound
groups of star-forming gas particles (i.e., gas particles with
densities above the threshold for star formation) and star particles.
We initially associate each {\sc skid}-identified galaxy with a dark
matter halo by using a spherical overdensity algorithm, defining the
virial radius as the radius enclosing a mean density given by \citet{kitayama96}.
Overlapping halos are then grouped together so
that, by construction, every halo in our final catalog has a central
galaxy (the most massive galaxy) and a number of satellite galaxies.
Beginning at $z = 2$, we reconstruct the evolution of each individual
galaxy back in time by identifying each galaxy's most massive
progenitor at all redshifts.  All galaxies presented here are
centrals.

When calculating morphological and kinematic properties of each galaxy
as a function of time, we take the position of the most bound gas
particle as the center of the galaxy, which corresponds to the
position of the black hole.  This results in a physically more meaningful and
a significantly more stable definition compared to the galaxy center
computed by {\sc skid}, especially during close galaxy encounters and
galaxy mergers.  Since we are interested in the evolution of galaxies
and black holes over cosmological timescales, we smooth the time evolution of
all physical quantities and extract their main evolutionary features
by calculating averages over time intervals of 50\,Myr (except
otherwise noted).  Finally, we impose a resolution limit on our sample
of galaxies by requiring that they contain at least 100 gas particles
and 100 star particles within the inner kiloparsec.  In this way, we
ensure that the morphology and the physical conditions at the centers
of galaxies can be accurately characterized for the purposes of this
work.  Therefore, lower mass galaxies take part of the analyses and
results only until they are resolved at later times.


\section{Galaxy Properties}\label{sec:galpro}

\begin{figure}[t]
\epsscale{1.0}
\plotone{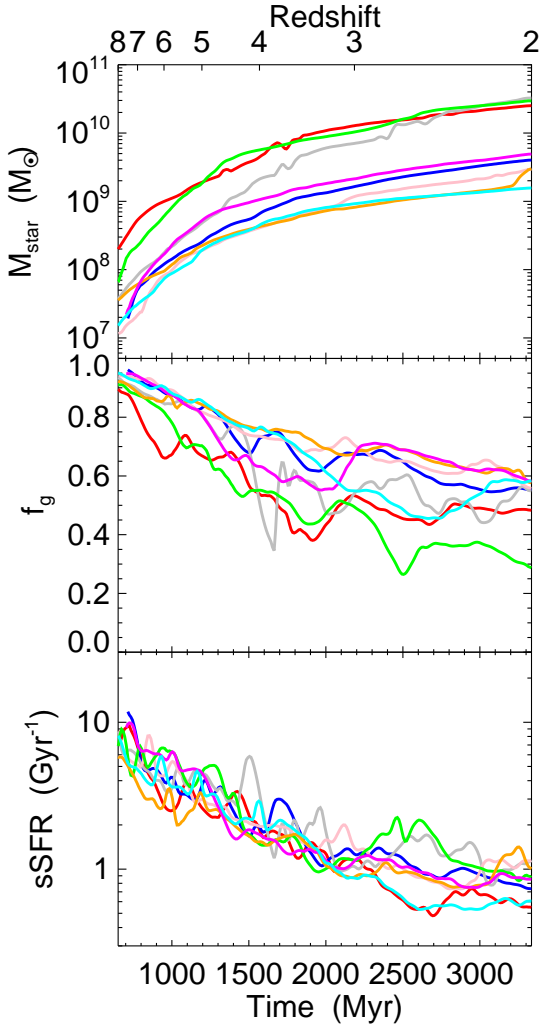}
\caption{{\it From top to bottom}: stellar mass, gas fraction, 
and specific SFR as a function of time from $z = 8$ to
$z = 2$ for all simulated galaxies.  Each color corresponds to a
different galaxy.}
\label{fig:evol}
\end{figure}

Our zoom simulations reproduce many characteristic features of
observed $z \approx 2$ star-forming galaxies.  The properties of these
galaxies have been analyzed in detail in a separate paper
\citep{ang12}, where we show that strong stellar winds are required in
order to maintain high gas fractions, redistribute star-forming gas
over larger scales, and increase the velocity dispersion of simulated
galaxies, in better agreement with the large, extended,
rotation-dominated yet turbulent star-forming disks of the SINS survey
\citep{for09}.

Here, we provide a general description of all simulated galaxies and
then we focus on the physical conditions at their centers, since they
eventually determine the accretion rates onto the central black hole.

\subsection{Global Properties}

Figure~\ref{fig:snaps} gives a visual impression of our galaxy sample
by showing the distribution of the star-forming gas and stellar
contents at $z = 2$.  For most galaxies, the star-forming gas lies
primarily in a thick, rotationally supported disk, with sizes ranging
from $\sim 5$\, to $\sim 25$\,kpc.  A wide range of morphologies can
be identified, including prominent spiral arms (galaxy g54), a very
compact gas distribution (galaxy g2403), a pair of interacting
galaxies (g2438), and a turbulent, highly disturbed galaxy (g222).
The projected stellar distributions also reveal a wide variety of
galaxy sizes and morphologies.  Most galaxies appear to have a large-scale stellar disk surrounded by a significant spherical component.

At $z = 2$, our simulated galaxies span about an order of magnitude in
both stellar mass (from $\sim 1.6 \times 10^{9}$ to $\sim 3.3 \times
10^{10}$\,\Msun) and SFR (from $\sim 1$ to 40\,\Msun\,yr$^{-1}$).
Figure~\ref{fig:evol} shows the evolution of individual galaxies from
$z = 8$ to 2 in terms of their stellar mass, gas fraction, and
specific SFR.  Major mergers (identified as abrupt changes in the
stellar mass of galaxies) can have a significant impact on the
dynamical properties of galaxies but represent only a small fraction
of the total mass growth down to $z = 2$.  Specific SFRs follow a
common trend for all galaxies, decreasing from $\sim 10$\,Gyr$^{-1}$
at $z = 8$ down to $\sim 1$\,Gyr$^{-1}$ at $z = 2$, but their
individual star formation histories are clearly different.  Gas
fractions evolve relatively slowly from nearly unity at $z = 8$ down
to $\sim 0.5$ at $z = 2$, due to the early suppression of star
formation by momentum-driven winds and the recycling of gas at later
times.  Overall, our simulated galaxies are characterized by a wide
range of masses, morphologies, and star formation histories and are
representative of ``normal" $z = 2$ systems \citep{ang12}.

\begin{figure*}[t]
\begin{center}
\includegraphics[scale=0.9]{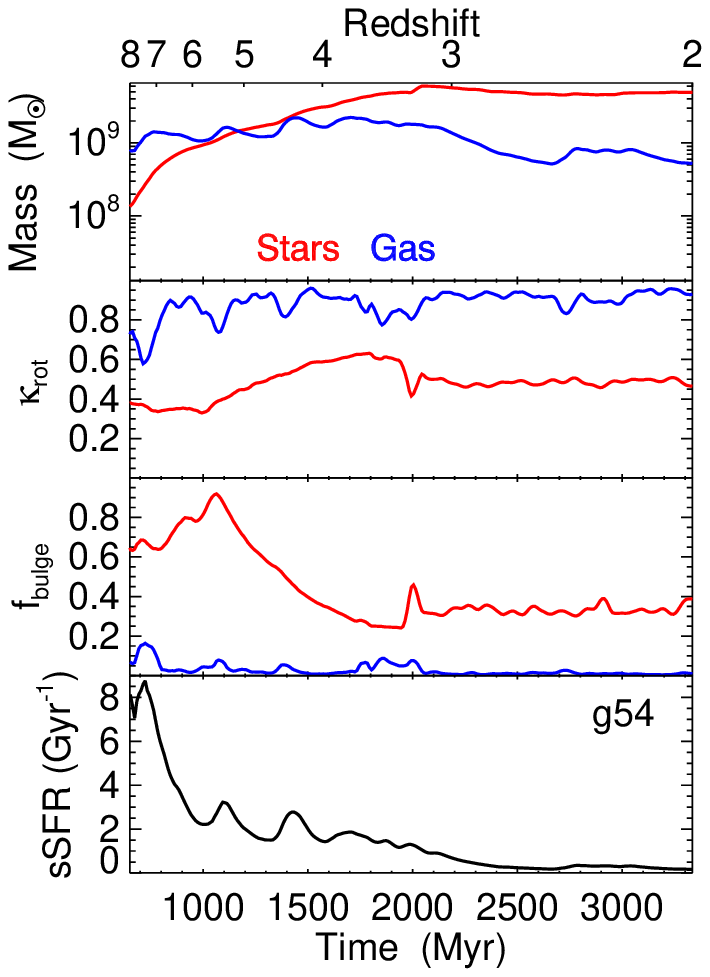}
\includegraphics[scale=0.9]{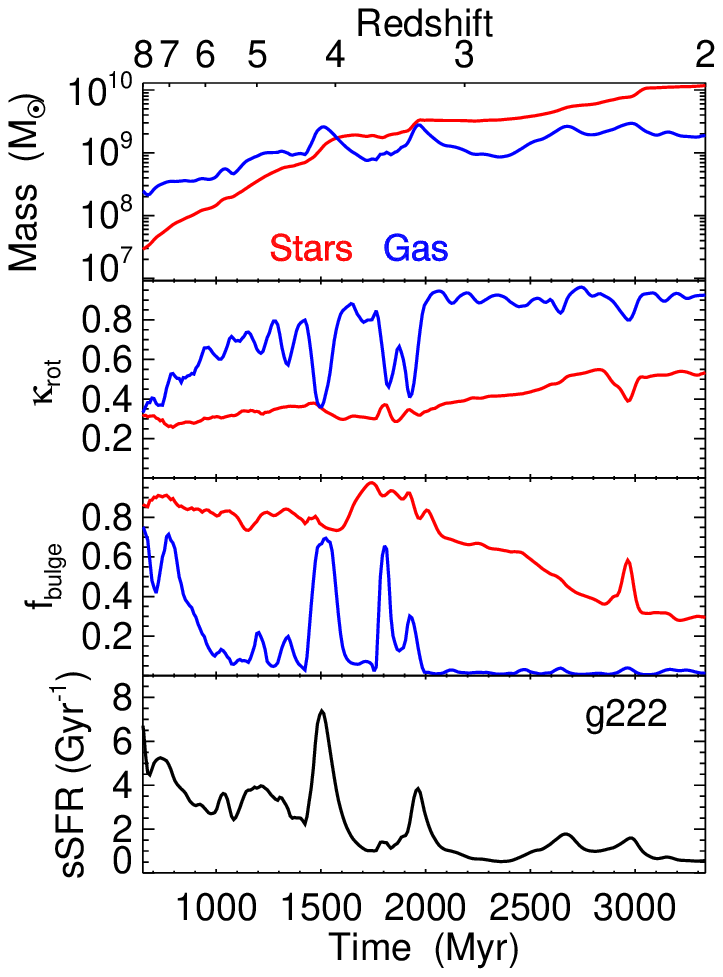}
\end{center}
\caption{Evolution of galaxy properties evaluated within the inner 
{\rm kpc} for the stellar (red line) and gas (blue line) components of
galaxies g54 ({\it left}) and g222 ({\it right}), two of the most
massive galaxies in the sample.  {\it From top to bottom}: total mass,
fraction of kinetic energy in ordered rotation (calculated with
respect to the total angular momentum of the galaxy), bulge mass
fraction (mass fraction in a spherical component calculated from
kinematic decomposition), and specific SFR.}
\label{fig:evolR0}
\end{figure*}

\subsection{The Central Kiloparsec}\label{sec:kpc}

In order to investigate the dominant processes governing black hole growth, it
is crucial to understand the physical conditions at the inner regions
of galaxies, how they evolve in time, and what the main evolutionary
triggers are.  Here, we evaluate the masses and morphologies of the
stellar and gas contents of galaxies as well as the specific SFRs
within a radial distance $R_{0} = 1$\,kpc from their centers.
Large-scale gravitational torques produced by galaxy mergers and
interactions are often invoked as triggers of star formation and AGN
activity \citep[e.g.,][]{spr05b}.  The identification of galaxy
mergers in cosmological simulations is, however, a non-trivial task in
which the simple working definition of ``galaxy" can result in a
rather different timing of merger events or the temporal
identification of close galaxy encounters as a merging system
\citep{gab11}.  For the purposes of this work, we simply associate
changes in galaxy properties with galaxy interactions and/or mergers
by visual identification from gas and stellar surface density plots
(similar to those in Figure~\ref{fig:snaps}) at the appropriate
snapshot times.

Figure~\ref{fig:evolR0} characterizes the masses and morphologies of
the stellar and gas contents of galaxies g54 and g222 within the inner
kiloparsec as a function of time.  In the remainder of this section,
we use these two galaxies to illustrate the analysis but we note that
our conclusions apply to all simulated galaxies.  Galaxy g54 undergoes
a relatively quiescent evolution below $z \approx 3$, after a merger
that results in a significant increase in the stellar mass within
$R_{0}$ ($\sim 2000$\,Myr after the big bang).  Galaxy g222 is located
in a denser environment and it undergoes a more violent evolution,
with mergers that result in the rapid increase of stellar and gas
masses within $R_{0}$ at times $\sim 1500$, 1900, 2700, and 3000\,Myr.

Here we parameterize the gas and stellar morphologies within the inner
kiloparsec of galaxies based on the amount of rotational support,
$\kappa_{\rm rot}$, and the bulge mass fraction, $f_{\rm bulge}$.  We
calculate $\kappa_{\rm rot}$ as the fraction of kinetic energy in
ordered rotation with respect to the total angular momentum within
$R_{0}$ \citep[e.g.,][]{sal10}.  Additionally, we estimate $f_{\rm
bulge}$ from a bulge-disk decomposition based on the full
three-dimensional kinematic information of particles of either the gas
or the stellar components.  We calculate the azimuthal velocity
($v_{\rm \phi}$) of each particle with respect to the direction of the
total angular momentum within $R_{0}$.  In the presence of a spherical
component supported by random motions, $v_{\rm \phi}$ is expected to
have a symmetric distribution with 50\,\% of the particles having
velocities $v_{\rm \phi} < 0$ on average.  Therefore, by adding up the
masses of particles with $v_{\rm \phi} < 0$ (and multiplying by 2) we
can get an estimate of the total mass in a spherical component.  This
will certainly underestimate $f_{\rm bulge}$ in the case of rotating
bulges but it is a reasonable approximation for the purposes of this
work.

Figure~\ref{fig:evolR0} shows that the gas component is more
rotationally supported than the stellar component at all times, since
the infalling gas can cool down and form a disk even if only a
fraction of the initial angular momentum is retained.  In the absence
of strong perturbations, the stellar $\kappa_{\rm rot}$ tends to
increase as the stars form from the gas disk (e.g., galaxy g54 from $z
= 6 \rightarrow 3$).  Galaxy interactions and mergers result in gas
inflows toward the center of galaxies and leave their imprint in
$\kappa_{\rm rot}$: an increase in the total gas mass within $R_{0}$
usually correlates with a decrease in $\kappa_{\rm rot}$ (e.g., galaxy
g222 at times $\sim 1500$ and 1900\,Myr).  As expected, $f_{\rm
bulge}$ anticorrelates with $\kappa_{\rm rot}$, since both quantities
are calculated based on particle kinematics.  We see that galaxy
interactions result in a temporary increase of the total gas mass as
well as the mass fraction in a spherical component.  Correspondingly,
the evolution of the specific SFR within the inner kiloparsec appears
to be modulated by galaxy interactions and mergers.  These events
generally trigger temporary increases in the specific SFR that
correlate with changes in the gas morphology and kinematics.  The
morphology of the stellar component can be affected by this enhanced
star formation activity, since stars forming from low angular momentum
gas after a merger can result in the increase of the stellar bulge
component (e.g., galaxy g222 at $t \sim 1500$\,Myr).

In summary, global properties of simulated galaxies and their
evolution on cosmological timescales are dominated by smooth
accretion and wind-regulated star formation \citep{ang12}.  Despite
this, galaxy interactions and mergers can have a significant impact on
the morphology and star formation properties at the inner regions of
galaxies.  As we show in the next section, these events will have an
effect on the inferred central black hole accretion rates and the evolution of
AGN activity.

\begin{figure*}[tp]
\begin{center}
\includegraphics[scale=0.9]{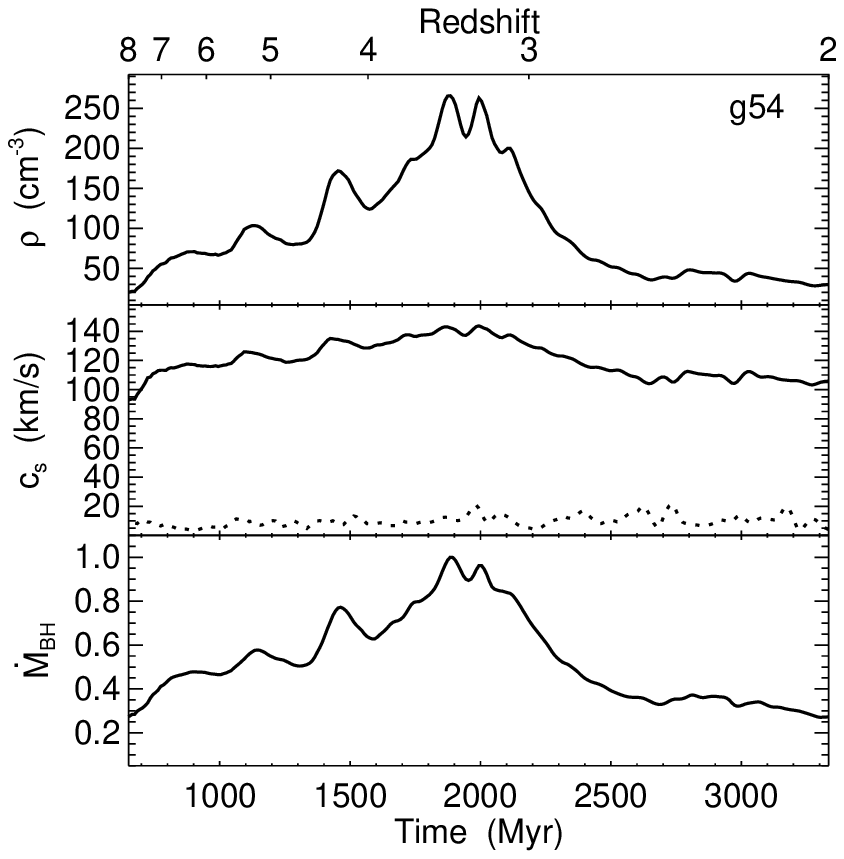}
\includegraphics[scale=0.9]{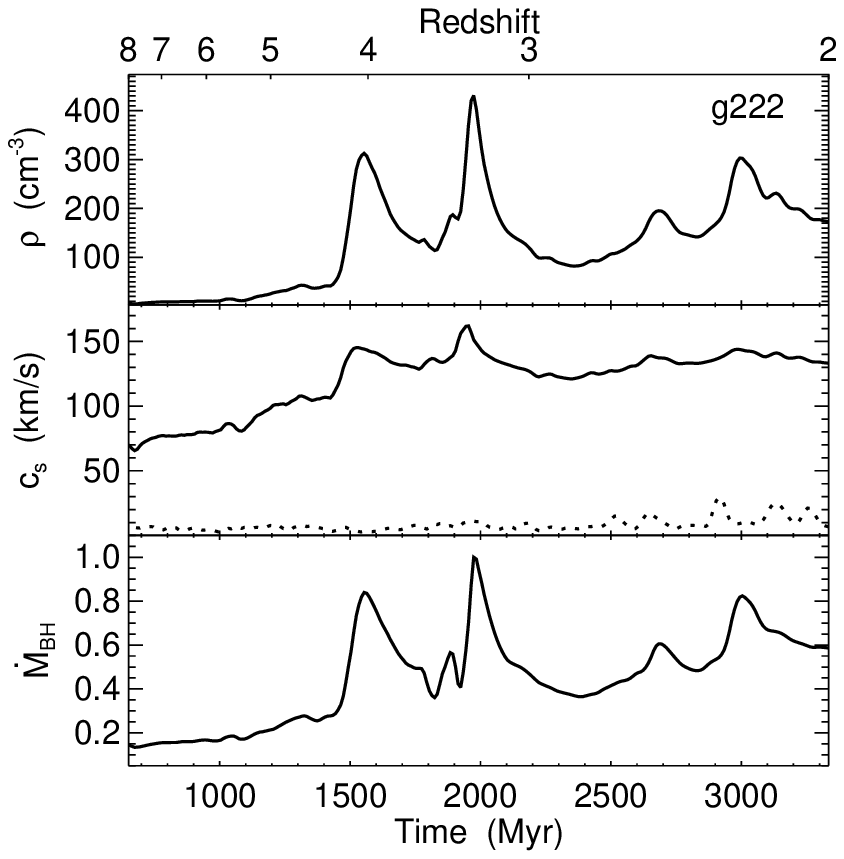}
\end{center}
\caption{Inferred Bondi rates for galaxies g54 ({\it left}) and 
g222 ({\it right}).  {\it Top}: gas density at the location of the
most bound gas particle.  {\it Middle}: sound speed at the location of
the most bound gas particle (solid line) and relative velocity with
respect to the surrounding gas (dotted line). {\it Bottom}: inferred
black hole accretion rates using the Bondi--Hoyle--Littleton parameterization
(Equation~(\ref{eq:bondi}) with $\alpha = 100$) and based on the
gas properties at the location of the most bound gas particle, where
we have used a constant black hole mass $M_{\rm BH} = 10^5$\,\Msun~at all times 
and normalized to the peak accretion rate.}
\label{fig:bondi}
\end{figure*}

\begin{figure*}[bp]
\begin{center}
\includegraphics[scale=0.9]{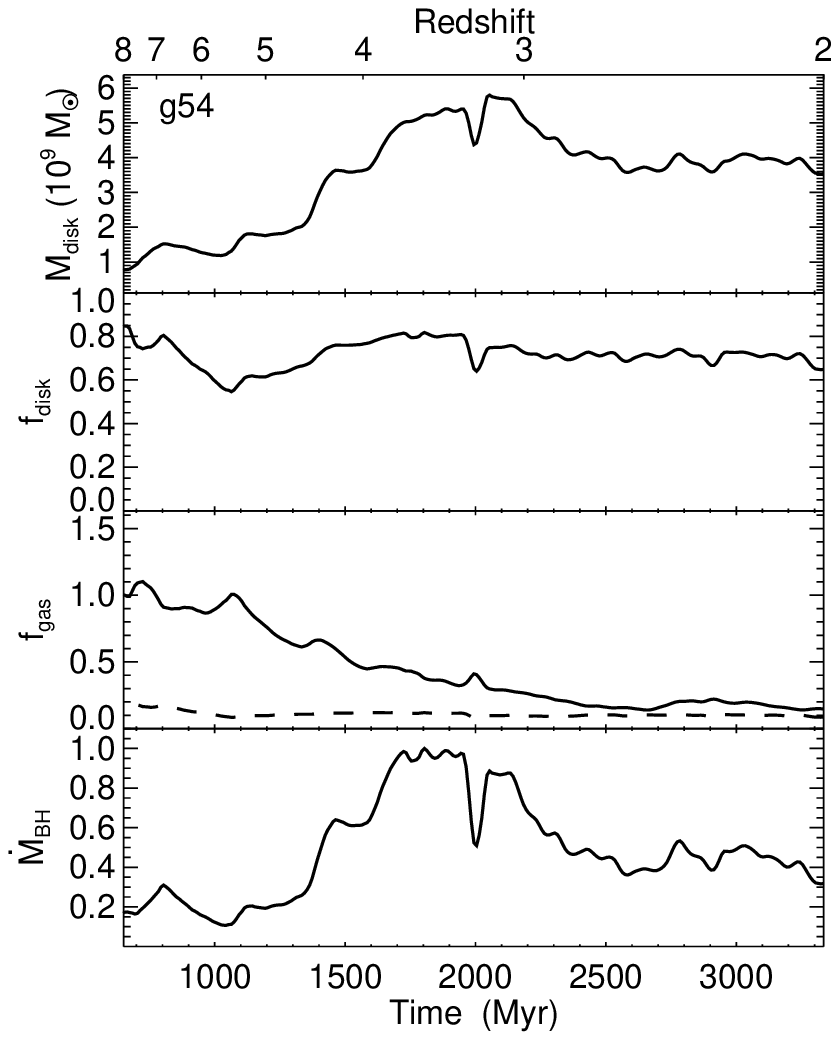}
\includegraphics[scale=0.9]{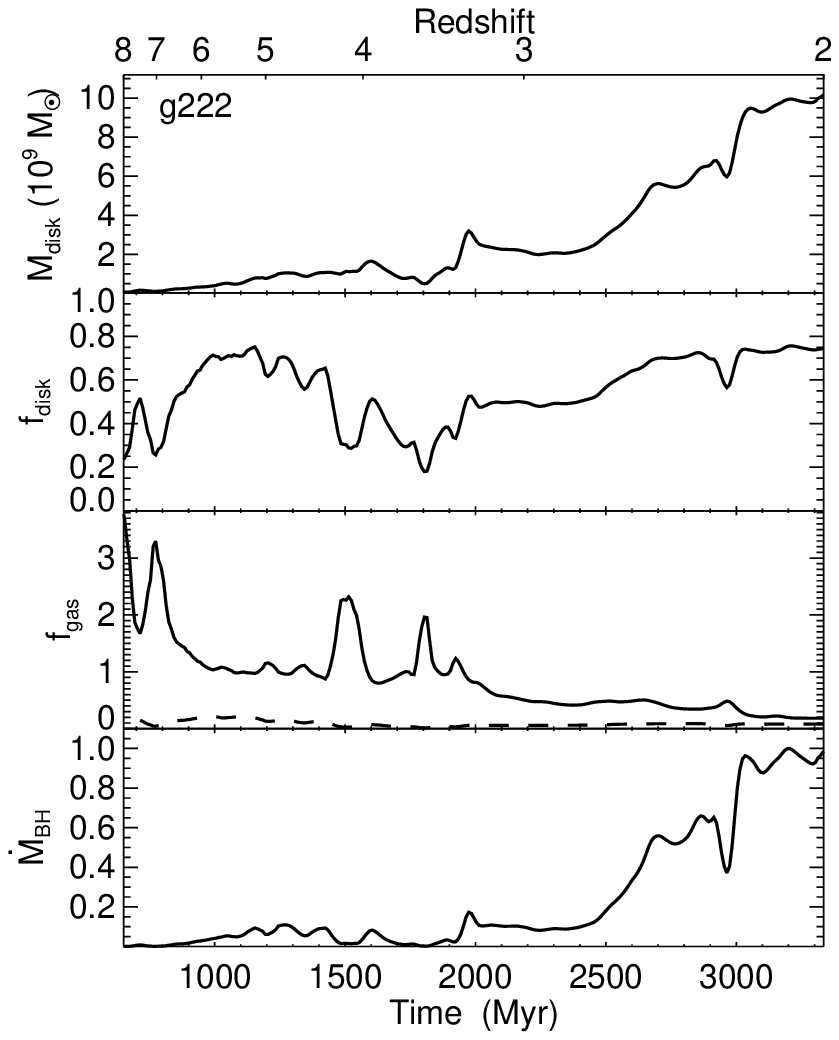}
\end{center}
\caption{Inferred gravitational torque rates for galaxies g54 
({\it left}) and g222 ({\it right}).  From top to bottom: (1) total (stellar and gas) disk mass evaluated within $R_{0} =
1$\,kpc, (2) total disk mass fraction within $R_{0}$, (3) ratio of gas mass to total (stellar and gas) disk mass evaluated
at $R_{0}$ (solid line), provided that $f_{\rm gas} \geq f_{0}$
(dashed line) inflow rates are not limited by gas supply, and (4) inferred black hole accretion rates using the analytic model of
\citet{hop11} (Equation~(\ref{eq:torque})), where we have used a
constant black hole mass $M_{\rm BH} = 10^5$\,\Msun~at all times and
normalized to the peak accretion rate.}
\label{fig:torque}
\end{figure*}


\section{Central Accretion Rates}\label{sec:acc}

We infer the accretion rates onto black holes located at the center of each
simulated galaxy based on two different accretion prescriptions. The
first is the spherical Bondi accretion, in which the mass accretion
rate is determined by the amount of matter captured gravitationally by
the black hole.  The second is the accretion rate driven by torques produced
by gravitational instabilities at different scales in the inner
galaxy.  We utilize the analytic model developed by \citet{hop11} to
calculate the latter accretion rate.  Since the evolution of the
central kiloparsec of galaxies has been fully characterized, we can
evaluate black hole accretion rates resulting from different physical
processes and investigate the connection of AGN activity and host
galaxy properties.  As we show below, the use of a particular black hole
accretion model in galaxy formation simulations can lead to rather
different predictions on the galaxy--AGN connection.
 
We choose the most bound particle in each galaxy to represent the
central black hole. However, in this set of simulations, we do not
self-consistently update the black hole mass and, therefore, we neglect the
gravitational influence of the black hole at the scales resolved in the
simulation.  The Bondi radius \citep{bon52} of a black hole of mass $M_{\rm
BH} = 10^{7}$\,\Msun, assuming a sound speed $c_{\rm s} = 30\,{\rm
km\,s^{-1}}$, is $r_{\rm B} = GM_{\rm BH} / c_{\rm s}^{2} \approx
50$\,pc, which is below the physical softening length in our
simulations for the redshift range of interest ($\epsilon \approx
75$--224\,pc at $z = 8$--2).  Furthermore, the typical masses of
central black holes that we infer for our simulated galaxies (see
Section~\ref{sec:growth}) are, at most, comparable to the mass of a
few tens of gas particles and we typically resolve the inner
kiloparsec of galaxies with thousands of gas particles.  Therefore,
given the mass and force resolution in our simulations, we do not
expect the lack of gravitational force from the central black hole to have a
significant effect on accretion rate estimates.  It can be argued,
however, that a more critical issue is the lack of black hole feedback in our
simulations.  We will address the issue of feedback in
Section~\ref{sec:growth}.

\subsection{Bondi--Hoyle--Littleton Parameterization}

The Bondi model \citep{hoy39,bon44,bon52} is the most widely used
prescription for black hole growth in galaxy formation simulations
\citep[e.g.,][]{spr05b,dimat08}.  For a black hole of mass $M_{\rm BH}$,
moving at velocity $v$ relative to a uniform distribution of gas with
density $\rho$ and sound speed $c_{\rm s}$, the Bondi rate is given by
\begin{equation}\label{eq:bondi}
\dot{M}_{\rm Bondi} = \alpha \, \frac{4\pi \, G^{2} \, 
M_{\rm BH}^{2} \, \rho}{(c_{\rm s}^{2}+v^{2})^{3/2}},
\end{equation}
where $\alpha$ is a dimensionless parameter which is usually added to
boost accretion rates and partially compensate for the relatively high
mean gas temperatures resulting from the multi-phase sub-grid model of
star formation and/or the lack of the spatial resolution required to
resolve the Bondi radius \citep{boo09,joh09}. The choice of this
parameter, together with the initial black hole mass, can have a significant
effect on the early growth of black holes.  Here we use a constant value
$\alpha = 100$, similar to many previous studies
\citep[e.g.,][]{spr05b}.  For comparison, we also explore the
functional form introduced by \citet{boo09}, where $\alpha \propto
\rho^{2}$ for gas at densities above the threshold for star formation
($n > 0.13$\,cm$^{-3}$ in our simulations) and $\alpha = 1$ for lower
density, single-phase gas.  We note that further modifications to the
Bondi parameterization have been proposed for the case of efficient
cooling and significant contribution of the surrounding halo to the
total gravitational potential \citep{hobb12}.
Making the reasonable assumption that the black hole is located at the center
of the potential well, we can get an estimate of the Bondi rate based
on the properties (gas density and sound speed) of the most bound gas
particle at the center of each simulated galaxy.

Figure~\ref{fig:bondi} shows the evolution of the density and sound
speed at the location of the black hole in galaxies g54 and g222.  Beginning
at $z = 8$, the central density of galaxy g54 increases by over two
orders of magnitude up to $\sim 250$\,cm$^{-3}$ at the end of the last
merger ($t \sim 2000$\,Myr), and then decreases rapidly down to $\sim
30$\,cm$^{-3}$ at $z = 2$.  Prior to the last merger, the accretion of
low-mass gas-rich satellites results in temporary increases of the
central density that correlate with morphological changes and
increases in specific SFRs as seen in Figure~\ref{fig:evolR0}.  The
effects of mergers in the central density of galaxy g222 are more
evident, with significant density peaks (up to $\sim 400$\,cm$^{-3}$)
clearly correlating with changes in the specific SFR within the inner
kiloparsec.

We evaluate the relative velocity of the surrounding gas ($v$) as a
SPH-kernel weighted averaged with respect to the most bound gas
particle, resulting in significantly lower values compared to the
typical sound speed (Figure~\ref{fig:bondi}).  For both galaxies, the
evolution of the central sound speed resembles that of the density but
with less than a factor of three variation during the simulation.  We note
that the large values of sound speed shown here are due to the
effective equation of state resulting from the sub-grid prescription
of star formation.  This leads to a significant suppression of Bondi
rates that is partially compensated by the addition of the boost
factor $\alpha$ in Equation~(\ref{eq:bondi}) \citep[see][for an
alternative approach]{pel07}.

The Bondi rates inferred for black holes located at the centers of galaxies
g54 and g222 are shown in Figure~\ref{fig:bondi}.  Since the goal
here is to identify the physical drivers of black hole accretion, we simply
evaluate Equation~(\ref{eq:bondi}) for a nominal, constant black hole
mass $M_{\rm BH} = 10^5$\,\Msun~at all times and show normalized Bondi
rates relative to the peak value for each galaxy.  This
simplification allows us to avoid making assumptions about the actual
growth of black holes at this stage and to focus only on how the evolution of
the host galaxy relates to relative changes in black hole accretion according
to the Bondi model.  Figure~\ref{fig:bondi} shows that the evolution
of the normalized Bondi rates resembles that of the central density
and it is therefore imprinted with the effects of galaxy mergers.
Since mergers invariably cause an increase in the central density of
galaxies, the Bondi model predicts a direct connection between AGN
activity and galaxy mergers, as has been found in many previous
simulations \citep[e.g.,][]{dimat05,spr05b}. Moreover, since star
formation is also proportional to the gas density
\citep[the sub-grid model is tuned to match the observed relation
of][]{ken98}, the Bondi model provides support for a merger-driven
scenario for the origin of starbursts and quasar phases \citep{hop06}.

\subsection{Accretion due to Gravitational Torques}
 
In any realistic flow with a non-zero net angular momentum, the gas is
expected to settle down onto a rotationally supported disk.  The rate
at which the gas is accreted depends on the rate at which its angular
momentum is removed.  Here, we evaluate black hole accretion rates based on
the analytic model of \citet{hop11} that accounts for the angular
momentum transport in the galactic disk.

The gravitational torque model predicts inflow rates at sub-parsec
scales as a function of physical quantities evaluated at scales that
can be resolved in the simulation.  Based on this prescription, the black hole
accretion rate (1) increases linearly with the total (stellar
and gas) disk mass, $M_{\rm disk}$, (2) depends strongly on the
total (stellar and gas) disk mass fraction, $f_{\rm disk}$, (3) increases with the ratio of gas mass to total disk mass, $f_{\rm
gas}$, and (4) depends very weakly on the black hole mass, $M_{\rm BH}$
\citep{hop11}:
\begin{align}\label{eq:torque}
& \dot{M}_{\rm Torque} \approx \alpha_{\rm T} \, f_{\rm disk}^{5/2} \times \left ( \frac{M_{\rm BH}}{10^{8}\,M_{\odot}} \right )^{1/6} \left ( \frac{M_{\rm disk}(R_{0})}{10^{9}\,M_{\odot}} \right )^{1} \nonumber \\
 & \times \left ( \frac{R_{0}}{100\,{\rm pc}} \right )^{-3/2}  \left (1 + \frac{f_{0}}{f_{\rm gas}} \right )^{-1} \, M_{\odot}\,{\rm yr^{-1}},
\end{align}
where
\begin{equation}
f_{0} \approx 0.31 \, f_{\rm disk}^{2} \, (M_{\rm disk}(R_{0})/10^{9}M_{\odot})^{-1/3},
\end{equation}
\begin{equation}
f_{\rm gas}(R_{0}) \equiv M_{\rm gas}(R_{0})/M_{\rm disk}(R_{0}),
\end{equation}
and $\alpha_{\rm T}$ is a normalization factor of order $\sim 1$--10
that parameterizes the dependence of inflow rates on star formation at
scales not resolved.  Here we use $\alpha_{\rm T} = 5$ \citep{hop11}
and evaluate all quantities within a radius $R_{0} = 1$\,kpc that is
well resolved in our simulations.  The total disk mass ($M_{\rm
disk}$) is calculated from kinematic decomposition (similar to $f_{\rm
bulge}$ in Section~\ref{sec:kpc}) and the total disk mass fraction is
calculated as $f_{\rm disk} = M_{\rm disk} / (M_{\rm
gas}(R_{0})+M_{\rm star}(R_{0}))$, with $M_{\rm gas}(R_{0})$ and
$M_{\rm star}(R_{0})$ the total gas and stellar masses within $R_{0}$,
respectively.

Figure~\ref{fig:torque} shows the evolution of $M_{\rm disk}$, $f_{\rm
disk}$, and $f_{\rm gas}$ for galaxies g54 and g222 as well as the
resulting black hole accretion rates according to the gravitational torque
model (evaluated for a constant black hole mass $M_{\rm BH} = 10^{5}$\,\Msun,
as before).  In our simulations, relative changes in the gravitational
torque rate primarily follow from the evolution of $M_{\rm disk}$ and
are, therefore, very sensitive to the morphology of the inner region
of the galaxy.  The ratio of the gas mass to the total disk mass,
$f_{\rm gas}$, tends to decrease with time as gas is transformed into
stars and the disk mass increases, except during galaxy interactions
that cause a temporary increase in gas mass and bulge fraction within
$R_{0}$ (e.g., galaxy g222 at $\sim 1500$ and 1900\,Myr).  Provided that
$f_{\rm gas} > f_{0}$, as it is the case in our simulations down to $z
= 2$, the gravitational torque rate is fairly insensitive to the gas
fraction, since densities at small scales are set by an equilibrium
between gas inflows and star formation \citep{hop11}.

The total disk mass within the inner kiloparsec of galaxy g54
increases by a factor of $\sim 5$ from $z = 6$ to $z = 3$.  Most of
the accreted gas settles onto a rotationally supported disk from which
stars form, resulting in a significant decrease in the stellar mass
fraction in a spherical component ($f_{\rm bulge}$;
Figure~\ref{fig:evolR0}).  Interestingly, galaxy mergers between $z
\approx 3$ and 6 seem to increase the disk mass and therefore the black hole
accretion rate, except for a temporary decrease in disk mass at the
end of the last merger ($t \sim 2000$\,Myr) that results in a
reduction of black hole accretion that lasts $\sim 100\,$Myr, until the disk
fraction rises again.

So far, we inferred accretion rates for a constant black hole mass 
at all times and it is, therefore, not possible to provide a direct 
comparison between the Bondi and gravitational torque models.
Despite this, the relative changes in black hole accretion as predicted by the
two models seem roughly consistent for galaxy g54, with significant
increases in black hole accretion rates correlating with galaxy mergers.
However, this scenario changes when looking at the evolution of galaxy
g222: mergers at times $t \sim 1500$ and 1900\,Myr cause a significant
increase in Bondi rates (due to the increase in central density;
Figure~\ref{fig:bondi}) but not in the gravitational torque rates.  In
this case, galaxy mergers inhibit the formation of a massive disk
component within the inner kiloparsec.  The bulge gas fraction
increases significantly for a period of $\sim 100$--200\,Myr after
each merger, from which stars quickly form without contributing to the
disk component (Figure~\ref{fig:evolR0}).  It is not until the end of
the second merger that the disk mass increases for both the gas and
stellar components and the black hole accretion rate increases according to
the gravitational torque model.  Interestingly, the last two mergers
at times $t \sim 2700$ and 3000\,Myr favor the rapid formation of a
massive disk component with the consequent increase in black hole accretion
rates.

In summary, galaxy mergers always cause an increase in the Bondi rate
due to the increase in density at the center of galaxies, but the
effects of mergers on the gravitational torque rates depend on whether
they cause an increase or a decrease in the disk mass fraction.
Large-scale gravitational torques produced by mergers have the
potential to remove angular momentum and drive gas inflows toward the
centers of galaxies, as we have seen for two of our simulated
galaxies.  However, if a merger results in the overall reduction of
the disk component, further gravitational instabilities in the disk
are suppressed, with the consequent reduction in gas inflow rates.
The gravitational torque model yields, therefore, a less direct
correspondence between major merger events and rapid phases of black hole
growth, which may help explain observational studies showing that
active galaxies do not preferentially show merger signatures at $z
\sim 2$ \citep{koc12,mull12b,scha12} and at lower redshifts
\citep[][]{gab09,cis11b,boe13}.

\begin{figure*}[tp]
\begin{center}
\includegraphics[scale=1.0]{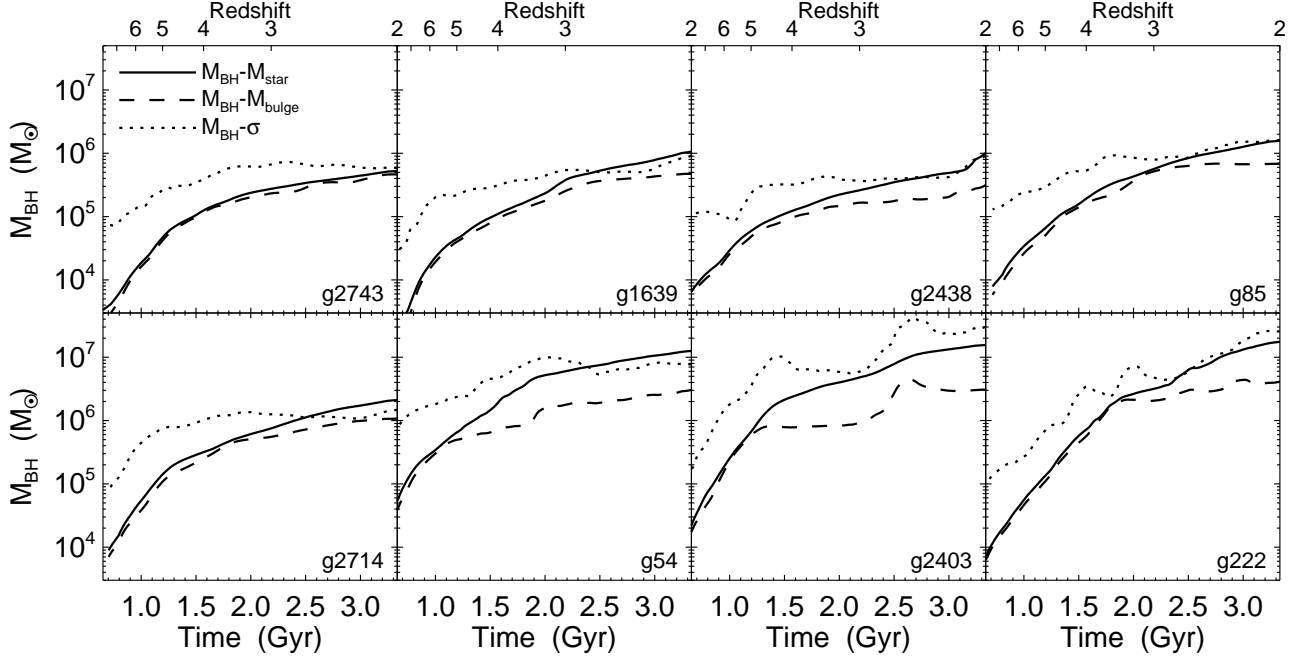}
\end{center}
\caption{Inferred masses of central black holes as a function of time for all 
simulated galaxies assuming that the observed $z = 0$ black hole--galaxy
correlations hold at all times.  For each galaxy, black hole masses are
calculated according to (1) the \Msig~relation \citep{trem02}
for the stellar velocity dispersion within the effective radius
(dotted line) and (2) the \Mbulge~relation \citep{har04} for
the total stellar mass (solid line) and the bulge mass (dashed line)
within the effective radius.  For each galaxy, black hole masses have been
averaged over timescales of $\sim 200$\,Myr.}
\label{fig:msig}
\end{figure*}

\begin{figure*}[bp]
\begin{center}
\includegraphics[scale=1.0]{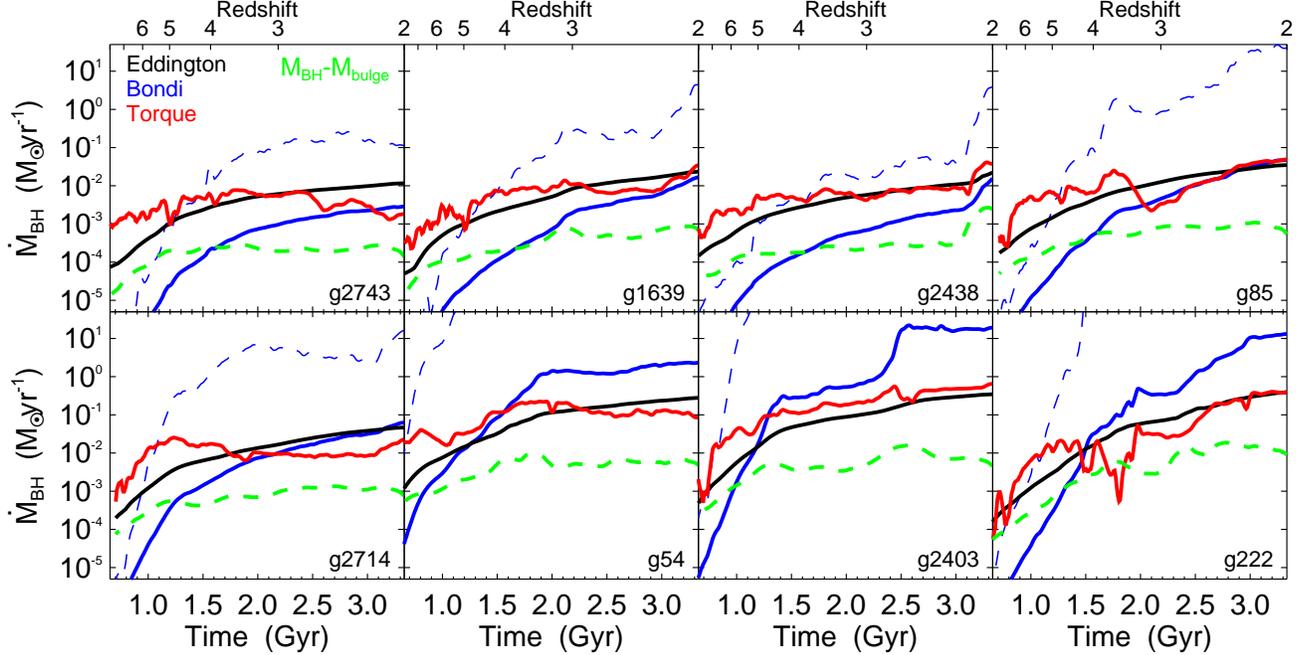}
\end{center}
\caption{Inferred accretion rates as a function of time for a black hole located 
at the center of each simulated galaxy.  Black hole masses are taken from the
\Mbulge~relation for the total stellar mass within the effective radius
 at all times (Figure~\ref{fig:msig}, solid line) and
accretion rates are calculated based on the properties of the host
galaxy over time, according to (1) the
Bondi--Hoyle--Littleton parameterization with $\alpha = 100$ (solid blue
line) and for the density-dependent $\alpha$ introduced by
\citet[][dashed blue line]{boo09}, (2) the gravitational torque
model of \citet[][red line]{hop11}, and (3) the Eddington rate
(black line).  Dashed green lines correspond to the actual accretion
rates required for black holes to grow according to the \Mbulge~relation for
each galaxy at all times (i.e., the time derivative of black hole masses shown
in Figure~\ref{fig:msig}).}
\label{fig:mdot}
\end{figure*}


\section{Constraints on Black Hole Growth}\label{sec:growth}

Accretion rates as predicted by the Bondi and gravitational torque
models depend on the effects of galaxy interactions and mergers on the
evolution of central galaxies.  More importantly, these two models
have a very different dependence on black hole mass.  Thus, in order to
provide a more meaningful comparison between these models and get a
better intuition on the physical processes governing cosmological black hole
growth they should be compared for a black hole that is growing with time in
some consistent way.

If we assume that the observed $z = 0$ black hole--galaxy correlations are
universal and do not depend strongly on redshift, then we can get a
reasonable estimate of black hole masses in terms of the properties of each
simulated galaxy.  Recent observations of active galaxies indicate a
possible evolution of the \Mbulge~(and perhaps \Msig) relation with
redshift \citep[e.g.,][]{dec10,mer10}.  Therefore, black hole masses derived
from the black hole--galaxy correlations should be taken just as a convenient
parameterization as a function of time.  In this way, we can infer the
absolute instantaneous accretion rates predicted by the Bondi and
gravitational torque models and evaluate the implications of different
physical processes in the evolution of black holes over cosmic time.

Figure~\ref{fig:msig} shows the mass of the central black hole as a function
of time calculated according to the \Msig~relation \citep{trem02} and
the \Mbulge~relation \citep{har04} for all simulated galaxies.  Here,
we calculate the effective radius of each galaxy ($R_{\rm eff}$) as
the two-dimensional projected radius enclosing one-half of the total
stellar mass.  The galaxy velocity dispersion ($\sigma$) is, then,
evaluated as the one-dimensional stellar velocity dispersion within
$R_{\rm eff}$.  Both $R_{\rm eff}$ and $\sigma$ are averaged over 100
random lines of sight.  The bulge mass is calculated as the stellar
mass in a spherical component within $R_{\rm eff}$ (from kinematic
decomposition; see Section~\ref{sec:kpc}) and also taken as the total
stellar mass within $R_{\rm eff}$ for simplicity.  Black hole masses shown in
Figure~\ref{fig:msig} have been averaged over time intervals of $\sim
200$\,Myr.

Black hole masses derived from the \Msig~relation are particularly noisy even
after smoothing over long timescales due to the finite resolution and
the calculation of $\sigma$ from individual particle motions.  We note
that the stellar bulge masses inferred from kinematic decomposition
usually represent lower limits to the ``true" bulge mass and it is
therefore not surprising that black hole masses decrease with time according
to the \Mbulge~relation during certain periods of galaxy evolution.
Despite this and given the inherently complex evolution of simulated
galaxies, it is encouraging that the \Msig~and \Mbulge~relations
predict black hole masses that are largely consistent with each other.

We can now compare the Bondi and gravitational torque models over
cosmic time by re-evaluating Equations~(\ref{eq:bondi})
and~(\ref{eq:torque}) for a black hole mass growing according to the black hole--galaxy
correlations.  Figure~\ref{fig:mdot} shows the evolution of the Bondi
and gravitational torque rates, as well as the corresponding Eddington
rates, for all simulated galaxies.  For simplicity, black hole masses have
been taken from the \Mbulge~relation for the total stellar mass within
$R_{\rm eff}$ (solid lines in Figure~\ref{fig:msig}).  We note that black hole
masses are fixed by the stellar mass of the parent galaxy over time
and therefore the inferred Bondi, gravitational torque, and Eddington
rates shown here represent instantaneous accretion rates at a given
time and do not reflect the actual growth of black holes according to each
model.

Some general trends can be identified in Figure~\ref{fig:mdot}.  At
very early times, black holes are small (with masses in the range $\sim
10^3$--$10^5$\,\Msun~according to the \Mbulge~relation of simulated
galaxies) and this results in a significant suppression of Bondi rates
due to the strong dependence on black hole mass ($\propto M_{\rm BH}^{2}$).
The Eddington rate increases linearly with black hole mass and, therefore, the
inferred instantaneous Bondi rate may eventually become significantly
higher than the corresponding Eddington limit for sufficiently massive
black holes.  In contrast, the gravitational torque model has a very weak
dependence on black hole mass ($\propto M_{\rm BH}^{1/6}$) and predicts gas
inflows from galactic scales to sub-parsec scales that can exceed the
Eddington limit by an order of magnitude at very early epochs.  The
gravitational torque rate also increases with black hole mass but is
significantly more sensitive to the evolution of galaxy properties.
According to this model, black holes would experience phases of both
super-Eddington and sub-Eddington feeding from $z = 8$ to $z = 2$.

\begin{figure*}
\begin{center}
\includegraphics[scale=0.8]{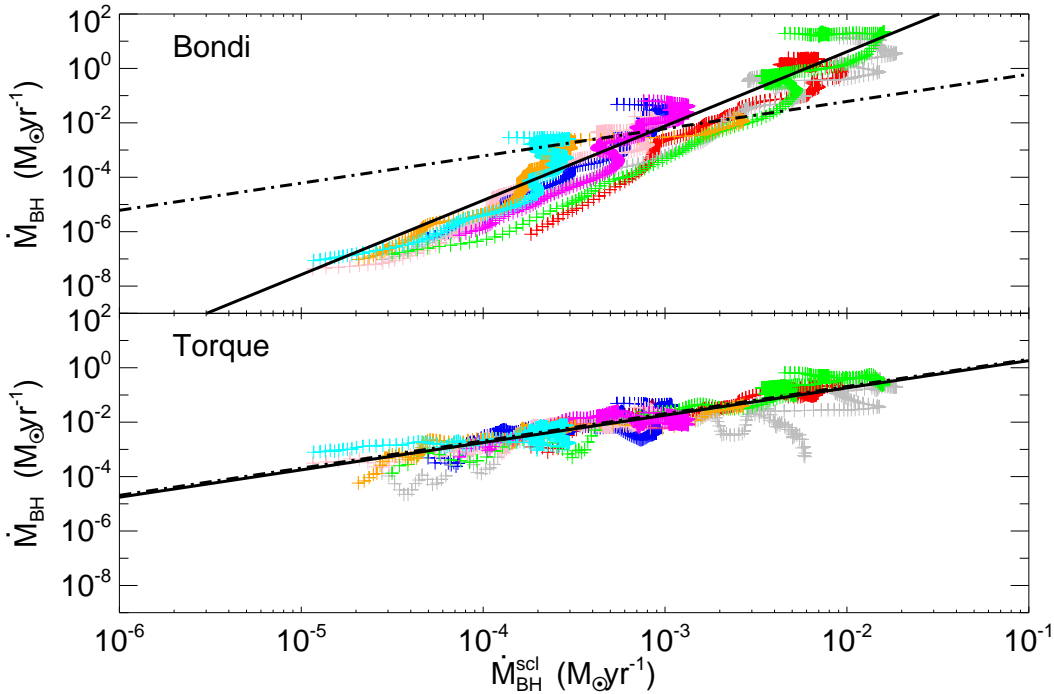}
\end{center}
\caption{Bondi rates with $\alpha = 100$ ({\it top}) and 
gravitational torque rates ({\it bottom}) as a function of the
accretion rate required for black holes to grow according to the
\Mbulge~relation for each galaxy at all times ($\dot{M}_{\rm scl}$).
Each color corresponds to a different galaxy as in
Figure~\ref{fig:evol}.  Solid lines show the best power-law fit for
each model and dash-dotted lines correspond to a linear relation
with constant slope given by $median(
\dot{M}_{\rm Bondi} / \dot{M}_{\rm scl})$ ({\it top}) and $median(
\dot{M}_{\rm Torque} / \dot{M}_{\rm scl})$ ({\it bottom}) for all
galaxies at all times.}
\label{fig:MdotScatt}
\end{figure*}

The dependence of accretion models on black hole mass can have profound
consequences on the inferred black hole evolution.  If black hole growth is limited by
the gravitational capture of gas as in Bondi, black holes will accrete gas at
very low rates early on and may never reach the conditions for
critical Eddington growth \citep[e.g.,][]{pel07}.  The rapid early
growth of supermassive black holes implied by observations of $z > 6$ quasars
\citep{fan03,mor11}, then, requires the formation of massive black hole seeds
($\sim 10^5$\,\Msun) for which early accretion is less suppressed
\citep[e.g,][]{dimat08}.  Once black holes are massive enough, the Bondi model
results in accretion rates well above the Eddington rate and an
additional mechanism regulating black hole growth is required.

The black hole mass at which the Bondi rate transitions from being
sub-Eddington to super-Eddington depends on model parameters and
technical implementations \citep[see, e.g.,][]{boo09,cho12}.  We
illustrate this in Figure~\ref{fig:mdot} by showing how the
density-dependent boost factor $\alpha \propto \rho^{2}$ results in
accretion rates that may differ by several orders of magnitude
compared to the constant $\alpha = 100$ model, shifting the transition
from sub- to super-Eddington to earlier times and, therefore, to lower
black hole masses. Despite this, the strong dependence of Bondi rates on the black hole
mass has clear consequences at the low-mass and high-mass limits
regardless of the particular value of $\alpha$ or other implementation
details.

If black hole growth is limited primarily by the transport of angular momentum
in the galactic disk, early growth can be rapid, since the inflow
rates predicted by the gravitational torque model can be well above
the Eddington limit even for small initial black holes.  Gas inflows are
driven by global gravitational instabilities in the disk and therefore
do not depend strongly on black hole mass.  Thus, the gravitational torque
model may ease constraints on models of black hole seed formation
\citep{vol10}.

Figure~\ref{fig:mdot} also shows the actual accretion rates required
for black holes to grow according to the \Mbulge~relation for each galaxy at
all times ($\dot{M}_{\rm scl}$).  Here we evaluate $\dot{M}_{\rm scl}$
numerically as the time derivative of the black hole mass calculated from the
\Mbulge~relation for each galaxy as a function of time
(Figure~\ref{fig:msig}).  We find that the inferred Bondi rates are
significantly lower than $\dot{M}_{\rm scl}$ at very early times and
become significantly higher than $\dot{M}_{\rm scl}$ toward $z = 2$
for all simulated galaxies.  This is a consequence of the strong
dependence on black hole mass, as discussed above.  In contrast, the
gravitational torque rate seems to follow the same dependence as
$\dot{M}_{\rm scl}$, but it is about an order of magnitude higher at
all times.  If black hole growth is limited primarily by the transport of
angular momentum by gravitational instabilities, it is, therefore, not
expected that all of the mass that is fed into the accretion flow at
$\lesssim 0.01$\,pc scales finds its way down to the central black hole.  In
fact, a number of theoretical and observational studies have shown
that only a small fraction of the mass inflow is retained in the
accretion flow, with the rest lost to winds and outflows
\citep[e.g.,][]{dimat00,king13,yua12}.

We compare in Figure~\ref{fig:MdotScatt} the inferred Bondi rates
(with $\alpha = 100$) and gravitational torque rates to $\dot{M}_{\rm
scl}$ for all galaxies at all times.  Remarkably, a simple power law
can provide a reasonable fit for all simulated galaxies. Specifically,
we find
\begin{equation}\label{eq:bscl}
 \dot{M}_{\rm Bondi} = (6.07\pm0.06) \dot{M}_{\rm scl}^{(2.73\pm0.02)},  
 \end{equation}
 \begin{equation}\label{eq:tscl}
 \dot{M}_{\rm Torque} = (1.26\pm0.03) \dot{M}_{\rm scl}^{(1.004\pm0.010)}.  
 \end{equation}
 
We note that, in most cases, large deviations from the best fit are
due to the calculation of black hole masses (and therefore $\dot{M}_{\rm
scl}$) directly from the evolution of galaxy masses.  The large power
index ($\sim 2.7$) for the Bondi rate confirms the trends found in
Figure~\ref{fig:mdot} for all simulated galaxies: Bondi rates evolve
from being significantly lower than $\dot{M}_{\rm scl}$ at small black hole
masses to being significantly higher at large black hole masses.  That is, black holes
growing at the Bondi rate would either lie well below the
\Mbulge~relation at all times or would become overly massive by $z =
2$.  If the Bondi rate provides a good estimate of the true accretion
rate, the large power index in Equation~(\ref{eq:bscl}) suggests that
(1) the initial black hole mass and the boost factor $\alpha$ need to be
set to allow for early growth and (2) some form of feedback
that becomes more efficient at higher accretion rates is required.
Indeed, suitable choices of model parameters and the coupling of a
fraction of the accretion luminosity to the surrounding gas in the
form of thermal energy have been successful in reproducing the
observed black hole--galaxy correlations \citep{dimat05,rob06,hop07,boo09}.

The gravitational torque rates are, on average, directly proportional
to the accretion rates required by the \Mbulge~relation, with a power
index very close to unity (Equation~(\ref{eq:tscl})).  Indeed, a linear
relation $\dot{M}_{\rm Torque} = \epsilon_{\rm m}^{-1} \, \dot{M}_{\rm
scl}$ with constant slope given by $\epsilon_{\rm m}^{-1} = median(
\dot{M}_{\rm Torque} / \dot{M}_{\rm scl})$ for all galaxies at all
times provides a very reasonable fit to the numerical results
(Figure~\ref{fig:MdotScatt}).  Remarkably, a simple mass retention
rate in the accretion flow, $\epsilon_{\rm m} \approx 5$\,\%, seems
sufficient to bring the gravitational torque rates down to
$\dot{M}_{\rm scl}$ over a range of more than three orders of
magnitude in accretion rates, for all simulated galaxies, and at all
times.  Therefore, the additional mechanism required to regulate black hole
growth according to the \Mbulge~relation has to be roughly equally
efficient regardless of the accretion rate.  Since the radiative
luminosity is proportional to the black hole accretion rate, the gravitational
torque model seems to disfavor AGN feedback acting at galactic scales
as the primary mechanism regulating black hole growth.

\section{Torque-limited growth}\label{sec:newmod}

The transport of angular momentum in the galactic disk is a required
process for black hole growth regardless of the presence and effects of AGN
feedback.  Therefore, we can explore an alternative scenario in which
winds and outflows from the inner accretion disk do not affect the
accretion flow and simply result in the reduction of black hole accretion
rates by a constant mass retention rate $\epsilon_{\rm m}$.  The
gravitational torque model describes gas inflows from galactic scales
down to $< 0.1$\,pc scales and therefore it can be interpreted as the
mass feeding rate into the black hole accretion disk.  Some of this mass may
be lost by radiatively driven outflows and other processes
\citep[e.g.,][]{pro00,ohs05,yua12}, while a small fraction reaches the
central black hole.  In this scenario, the central black hole grows on average at a
fraction $\epsilon_{\rm m}$ of the large-scale gas inflow rate:
 
\begin{equation}\label{eq:corr}
dM_{\rm BH}/dt = \epsilon_{\rm m} \, \dot{M}_{\rm Torque}(t).
\end{equation} 

\begin{figure}
\epsscale{1.2}
\plotone{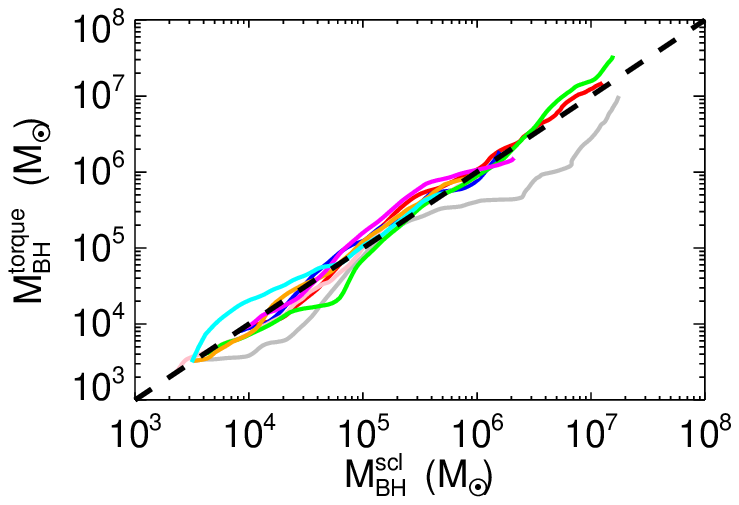}
\caption{Black hole mass calculated self-consistently from the gravitational 
torque rate corrected by the mass retention rate $\epsilon_{\rm m} =
0.05$ (by integration of Equation~(\ref{eq:corr})) as a function of black hole
mass derived from the \Mbulge~relation for each galaxy at all times.
Each color corresponds to a different galaxy as in
Figure~\ref{fig:evol}.}
\label{fig:NumBH}
\bigskip
\epsscale{1.2}
\plotone{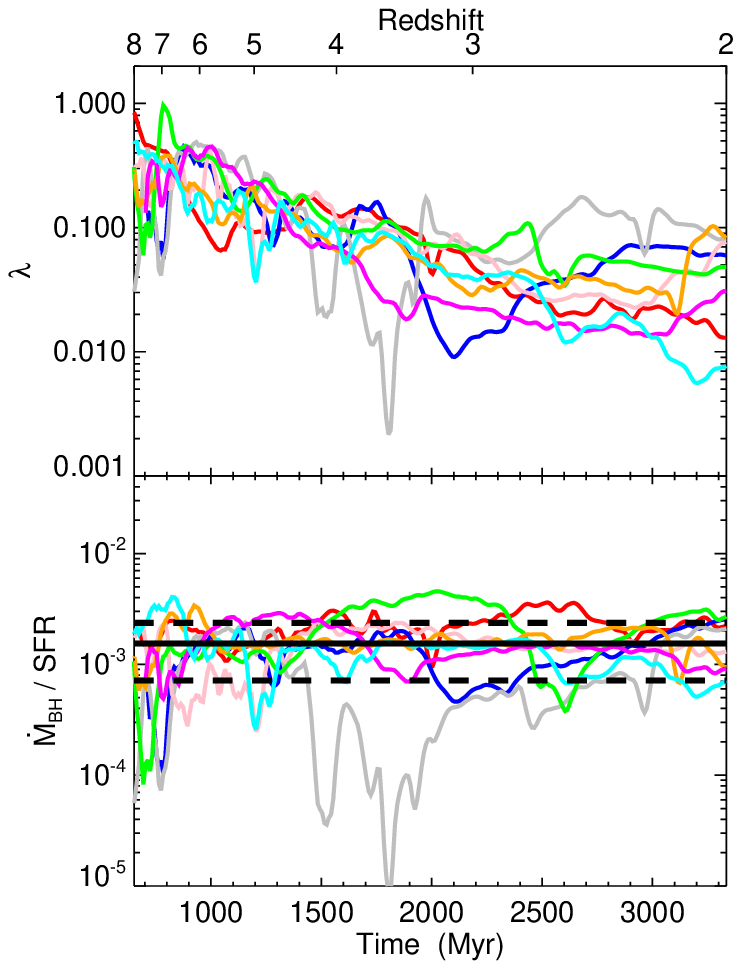}
\caption{{\it Top}: Eddington ratio (defined here as 
$\dot{M}_{\rm BH} / \dot{M}_{\rm Edd}$) for black holes growing
self-consistently at the gravitational torque rate corrected by the
mass retention rate $\epsilon_{\rm m} = 0.05$ (by integration of
Equation~(\ref{eq:corr})) for each galaxy as a function of time.  {\it
Bottom}: ratio of the black hole accretion rate to the total SFR within the
effective radius.  Black solid and dashed lines show a running average
of $\dot{M}_{\rm BH}/$SFR over all galaxies and all times and the
standard deviation, respectively.  Initial black hole masses are taken to be
consistent with the $z = 0$ \Mbulge~relation (Figure~\ref{fig:msig}).
Each color corresponds to a different galaxy as in
Figure~\ref{fig:evol}.}
\label{fig:EddRatio}
\end{figure}

Figure~\ref{fig:NumBH} shows the evolution of black hole masses predicted by
the gravitational torque model normalized by the mass retention rate
$\epsilon_{\rm m} \approx 0.05$ ($M_{\rm BH}^{\rm Torque}$), as a
function of the black hole mass obtained from the \Mbulge~relation for each
galaxy at each time step ($M_{\rm BH}^{\rm scl}$).  Here, we are
simply integrating Equation~(\ref{eq:corr}) for an initial black hole mass which
is consistent with the \Mbulge~relation for each galaxy.  We note that
$\dot{M}_{\rm Torque}(t)$ is self-consistently calculated based on the
morphological properties of each galaxy over time
(Equation~(\ref{eq:torque})) and updated with the appropriate black hole mass at
each time step, as given by Equation~(\ref{eq:corr}).  Remarkably, we
find that black holes evolve approximately along the $M_{\rm BH}^{\rm Torque}
= M_{\rm BH}^{\rm scl}$ line and therefore, remaining consistent with
the \Mbulge~relation.

In Section~\ref{sec:retrate}, we show that $\epsilon_{\rm m}$ sets the
normalization of the \Mbulge~relation.  We note, however, that the
gravitational torque model (Equation~(\ref{eq:torque})) contains a
normalization constant of order $\alpha_{\rm T} \approx 1$--10 that
parameterizes the effects of nuclear star formation on black hole accretion
\citep{hop11}.  Therefore, the normalization of the \Mbulge~relation
is actually determined by the product $\alpha_{\rm T}\,\epsilon_{\rm
m}$.  Given that plausible values of these two constants differ by
about two orders of magnitude, we choose to interpret them as
parameterizing different physical processes: $\alpha_{\rm T}$ controls
how much gas is converted into stars versus feeding the black hole accretion
disk \citep{hop11} and $\epsilon_{\rm m}$ controls what fraction of
the mass feeding the accretion disk is finally accreted by the black hole.  We
implicitly assume that the remaining mass, a fraction ($1 -
\epsilon_{\rm m}$) of the disk feeding rate, is expelled in the
polar direction without significantly halting further accretion.

We can now look back at the accretion histories of black holes growing
according to the gravitational torque model with the appropriate
normalization ($\epsilon_{\rm m} \approx 0.05$).
Figure~\ref{fig:EddRatio} shows the evolution of the inferred
Eddington ratios of black holes located at the center of all simulated
galaxies.  Despite the large scatter between galaxies and the high
temporal variability, there is a clear trend for Eddington ratios to
decrease from $\sim 0.1$--1 at $z = 8$ down to $\sim 0.01$--0.1 at $z
= 2$.  This is in agreement with recent observations of $z \gtrsim 4$
quasars showing that higher redshift black holes have lower masses but are
accreting at higher Eddington ratios compared to lower redshift black holes
\citep{dero11,tra11}.

Intriguingly, the inferred Eddington ratios evolve in a manner very
similar to the evolution of specific SFRs (Figure~\ref{fig:evol}):
both quantities seem to decrease by about a factor of 10 from $z = 8
\rightarrow 2$, on average.  In fact, Figure~\ref{fig:EddRatio} (lower
panel) shows that the ratio of the black hole accretion rate to the SFR within
the effective radius of galaxies is, on average, very close to the
ratio of black hole mass to stellar mass as inferred from the local
\Mbulge~relation:
\begin{equation}
\left \langle \frac{\dot{M}_{\rm BH}}{\rm SFR} \right \rangle \approx \, 
0.00154  \pm   0.0008,
\end{equation}
where the average (and standard deviation) has been calculated for all
galaxies and all times.  There is, of course, large scatter in
$\dot{M}_{\rm BH}/$SFR between different galaxies and evolution times,
but this ratio is, on average, remarkably flat in the redshift range
$z = 2$--8: the bulk of black hole and galaxy growth occurs in tandem, at
rates governed by cosmological infall and transport of angular
momentum in the galaxy disk, and with no feedback-mediated coupling
between black holes and parent galaxies.

The torque-limited model predicts, therefore, a natural connection
between AGN activity and SFR of galaxies on cosmological timescales,
in agreement with observations showing that the volume-averaged ratio
of black hole accretion rate to SFR agrees with the ratio of black hole mass to
stellar mass as inferred from the local \Mbulge~relation
\citep{heck04,zhe09}.  Furthermore, these results are consistent with
recent observational evidence for a constant black hole accretion to star
formation ratio independent of redshift since $z \sim 2$
\citep{raff11,mull12a}.  Since black holes and galaxies evolve approximately
along the scaling relations, the torque-limited model provides support
for a roughly constant black hole mass to stellar mass ratio independent of
redshift, consistent with the conclusions of \citet{jah09},
\citet{cis11a}, and \citet{mull12a} at redshifts $z \lesssim 2$.  The
AGN--SFR connection is, however, not direct in a galaxy-by-galaxy basis
and at all times (as evident from the scatter in
Figure~\ref{fig:EddRatio}, lower panel), which may help explain the
large scatter in observed AGN to star formation ratios in individual
systems \citep[e.g.,][]{raff11}.

Torque-limited growth is, therefore, a natural alternative to
feedback-regulated models and it is indeed consistent with our
``feedback-free" starting point, since black hole accretion rates have been
inferred based on galaxy properties from simulations with no AGN
feedback model (Section~\ref{sec:acc}).  AGN feedback may have a
significant impact on the host galaxy but it is not required for
regulating black hole growth.  Angular momentum transport (in the galaxy as
well as the accretion disk) together with competition with star
formation seem to be sufficient for black holes to grow according to the
scaling relations from early times down to $z = 2$.  
This is explored in more detail in the next section.


\begin{figure*}[t]
\begin{center}
\includegraphics[scale=1.1]{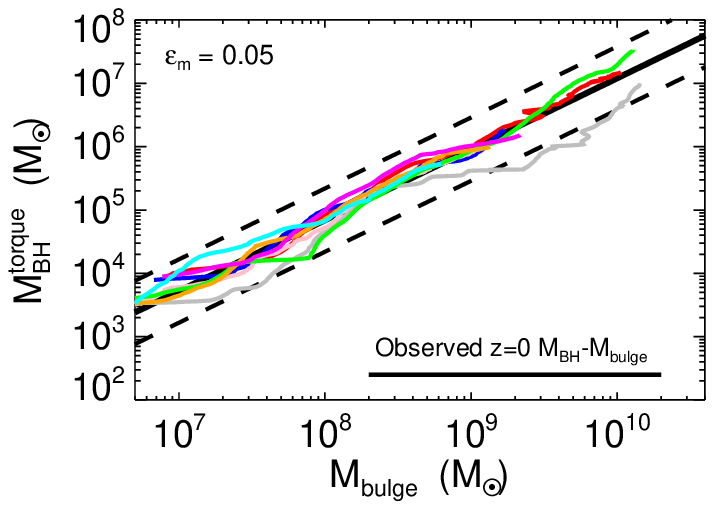}
\includegraphics[scale=1.1]{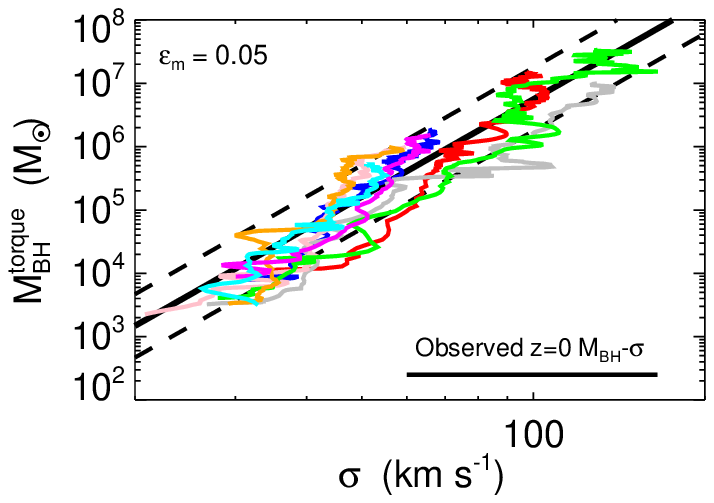}
\end{center}
\caption{\Mbulge~({\it left}) and \Msig~({\it right}) relations for black holes 
growing self-consistently at the gravitational torque rate corrected
by the mass retention rate $\epsilon_{\rm m} = 0.05$ (by integration
of Equation~(\ref{eq:corr})) for each galaxy at all times.  Initial black hole
masses are taken to be already consistent with the scaling relations.
Black solid lines show the \Mbulge~and \Msig~relations of
\citet{har04} and \citet{macc11}, respectively, and black dashed lines
indicate a 0.5 dex scatter in black hole mass for each relation.  The galaxy
velocity dispersion ($\sigma$) is calculated as the one-dimensional
stellar velocity dispersion within the effective radius ($R_{\rm
eff}$) averaged over 100 random lines of sight.  $M_{\rm bulge}$ is
taken as the total stellar mass within $R_{\rm eff}$.  Each color
corresponds to a different galaxy as in Figure~\ref{fig:evol}.}
\label{fig:NumRel}
\end{figure*}

\section{Black Hole--Galaxy Correlations}\label{sec:corr}

Figure~\ref{fig:NumRel} shows how galaxies and black holes evolve in the
\Mbulge~and \Msig~planes according to the torque-limited model
(Equation~(\ref{eq:corr})).  As we have seen in the previous section, a
simple normalization constant $\epsilon_{\rm m} \approx 0.05$ applied
to the gravitational torque model brings black holes and galaxies close to the
\Mbulge~relation of \citet{har04}.  Calculation of the galaxy stellar
velocity dispersion from particle motions results in higher scatter
but still shows that black holes growing according to Equation~(\ref{eq:corr})
are roughly consistent with the \Msig~relation of \citet{macc11}.

In Section~\ref{sec:growth} we have inferred ``instantaneous"
accretion rates according to the Bondi and gravitational torque models
by assuming that the observed \Mbulge~relation holds at all times.
This was a convenient way of parameterizing black hole mass as a function of
time for each galaxy, but there is a priori no reason to think that
galaxies and black holes should evolve together at all times.  The
gravitational torque rates inferred in this way were often of the
order of the Eddington limit (Figure~\ref{fig:mdot}) and therefore we
might be tempted to think that black holes growing at a fixed fraction of the
Eddington limit could also grow in a manner consistent with the
black hole--galaxy correlations.  However, once we release the assumption of
black hole--galaxy co-evolution (so that black hole masses are no longer given by the
\Mbulge~relation) and calculate black hole growth self-consistently over time,
we see that a simple normalization constant cannot compensate for the
exponential Eddington growth.  The fact that the torque-limited model
naturally leads to black holes growing on average according to the black hole--galaxy
correlations is a non-trivial consequence of the weak dependence on black hole
mass and the strong dependence on galaxy scale properties.

In the remainder of this section, we show how the inferred black hole--galaxy
correlations depend on model parameters and implementation details,
and discuss additional implications of the torque-limited model.

\subsection{Initial Black Hole Mass}\label{sec:ini}

\begin{figure*}
\begin{center}
\includegraphics[scale=1]{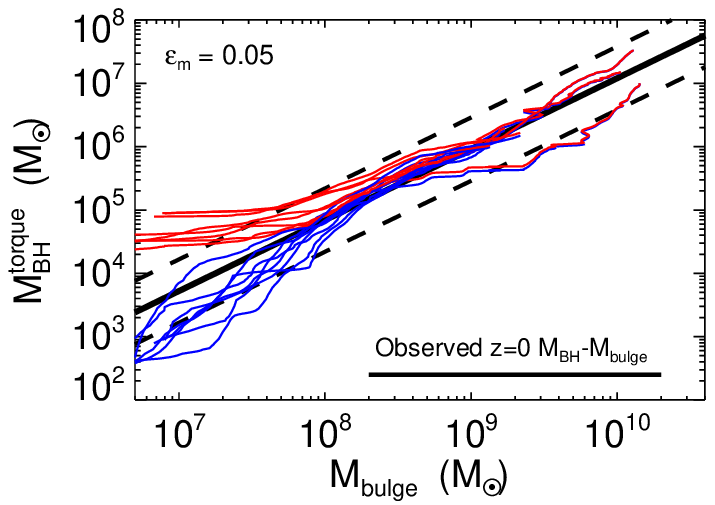}
\includegraphics[scale=1]{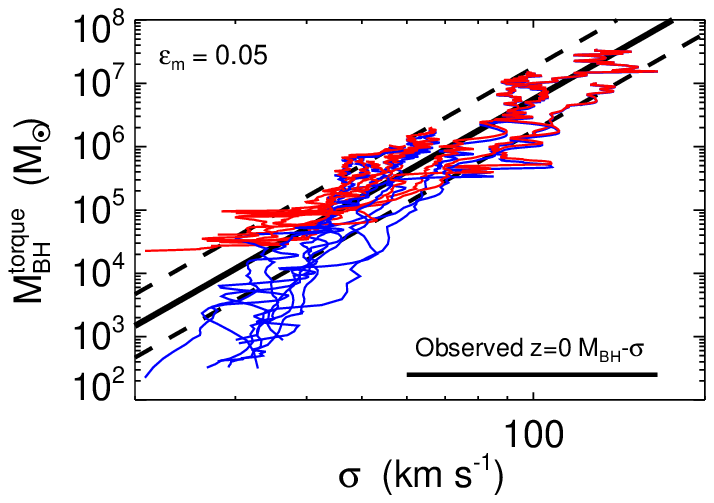}
\includegraphics[scale=1]{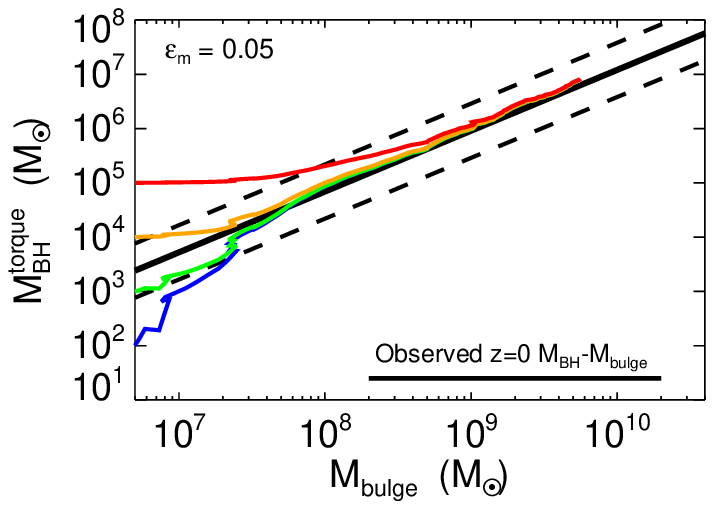}
\includegraphics[scale=1]{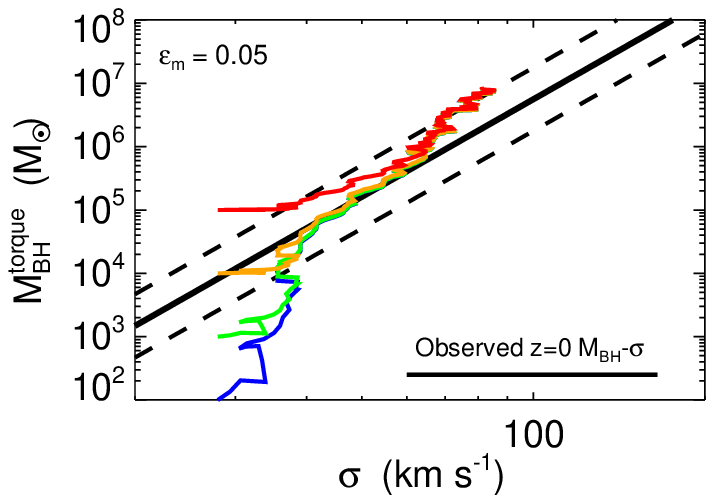}
\includegraphics[scale=1]{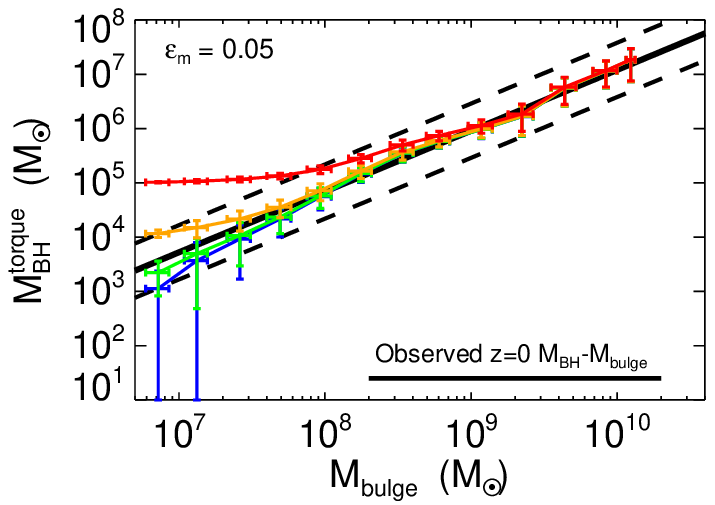}
\includegraphics[scale=1]{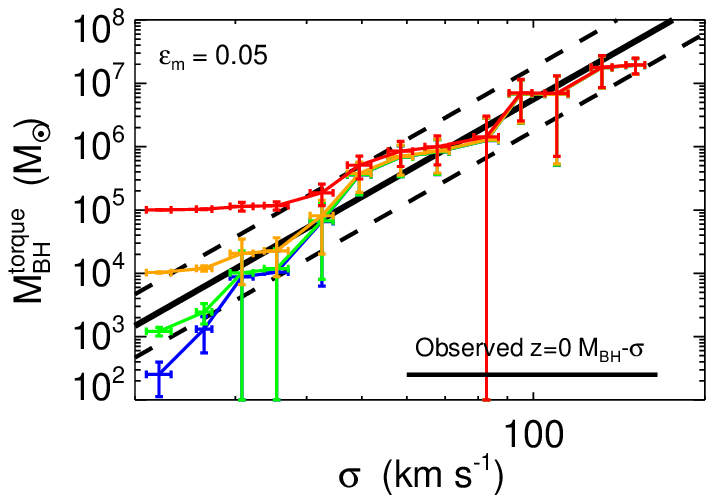}
\end{center}
\caption{\Mbulge~({\it left panels}) and \Msig~({\it right panels}) 
relations for black holes growing self-consistently at the gravitational
torque rate (Equation~(\ref{eq:corr}) with $\epsilon_{\rm m} = 0.05$)
for different initial black hole masses.  {\it Top panels}: the initial black hole
mass is taken to be either 10 times higher (red) or 10 times lower
(blue) compared to the scaling relation for each galaxy.  {\it Middle
panels}: black hole mass as a function of either bulge mass or velocity
dispersion for the ``average" galaxy (i.e., averaging $M_{\rm BH}$,
$M_{\rm bulge}$, and $\sigma$ over all galaxies at each time step) for
initial black hole masses ranging from $10^{2}$ to $10^{5}$\,\Msun~(from blue
to red). {\it Bottom panels}: average black hole mass within
logarithmically spaced bins in either bulge mass or velocity
dispersion, using all simulated galaxies at all time steps, and for
initial black hole masses taken from $10^{2}$ to $10^{5}$\,\Msun~(from blue to
red).  Error bars show the dispersion in each bin.  Black solid and
dashed lines correspond to the observed scaling relations
\citep{har04,macc11} and a 0.5 dex scatter in black hole mass, respectively.}
\label{fig:ini}
\end{figure*}

We have shown that provided the initial black hole mass lies approximately in
the \Mbulge~relation, the gravitational torque model scaled by a
simple constant factor predicts that black holes and galaxies evolve together
along the scaling relations.  However, black holes do not necessarily know
what is the correct initial mass for a given galaxy.  In
Figure~\ref{fig:ini}, we evaluate how changes in the initial
conditions affect black hole growth and the inferred \Mbulge~and
\Msig~relations.  First, we grow black holes with initial masses either a
factor of 10 above or below the scaling relations for each galaxy.  We
find that, regardless of the initial mass, black holes tend to evolve onto the
observed black hole--galaxy correlations.  In our simulations, accretion rates
are governed by the disk mass, with a very weak dependence on black hole mass,
and therefore black holes with different initial masses grow at comparatively
similar rates for a given galaxy.  Since most of the black hole mass comes
from accretion, which depends mostly on the evolution of the galaxy,
the initial conditions are erased and the black hole--galaxy correlations
are established.

This is also observed when the initial black hole masses are uncorrelated with
their parent galaxies.  Figure~\ref{fig:ini} (middle panels) shows the
\Mbulge~and \Msig~relations when initial black hole masses are taken to be the
same for all galaxies and ranging from $10^{2}$ to $10^{5}$\,\Msun.
For simplicity, we show the evolution of the ``average galaxy" that we
calculate by averaging black hole mass, bulge mass, and velocity dispersion
over all galaxies at each time step.  Provided we let black holes and galaxies
evolve for a sufficient amount of time, they all converge on average
toward the scaling relations.  Finally, we show in
Figure~\ref{fig:ini} (bottom panels) that this convergence toward the
scaling relations is also observed when we take black hole masses for all
galaxies at all times and average them within bins in either bulge
mass or velocity dispersion regardless of the evolution time of each
galaxy.

\begin{figure*}[t]
\begin{center}
\includegraphics[scale=1]{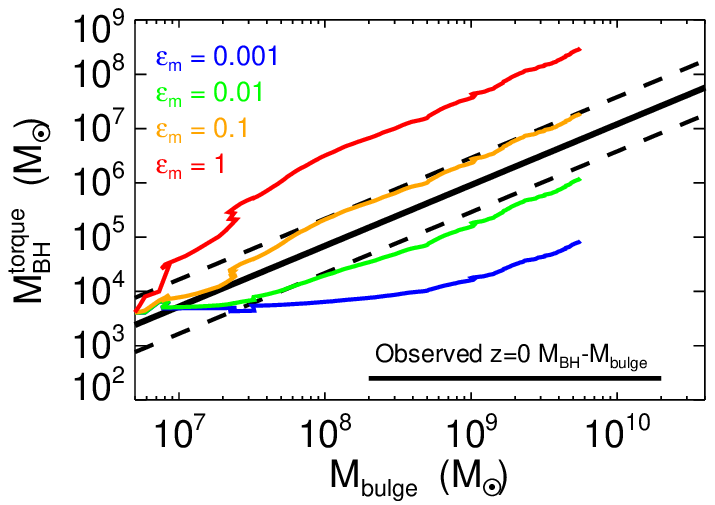}
\includegraphics[scale=1]{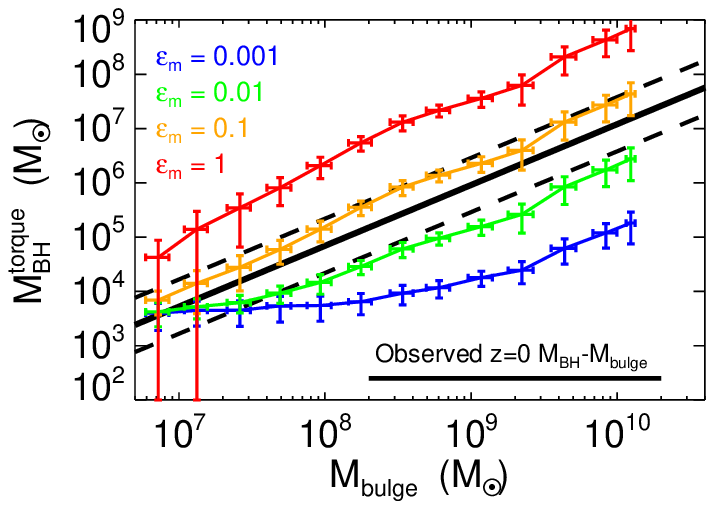}
\end{center}
\caption{Effects of varying the mass retention rate from 
$\epsilon_{\rm m} = 0.001$ to $\epsilon_{\rm m} = 1$ by factors of 10
(from blue to red) in the \Mbulge~relation.  {\it Left}: black hole mass as a
function of bulge mass averaged over all galaxies at each time step
for initial black hole masses consistent with the \Mbulge~relation.  {\it
Right}: average black hole mass within logarithmically spaced bins in bulge
mass for all simulated galaxies, at all time steps, and taking initial
black hole masses consistent with the \Mbulge~relation for each galaxy.  Black
solid and dashed lines correspond to the observed \Mbulge~relation
\citep{har04} and a 0.5 dex scatter in black hole mass, respectively.}
\label{fig:eff}
\end{figure*}

Black hole--galaxy correlations arising from the torque-limited model are
robust to changes in the initial mass of black holes and do not require
specific tuning of the mass retention rate $\epsilon_{\rm m}$.
Importantly, black holes with masses as low as 100\,\Msun~at $z = 8$ are able
to grow comparatively faster than higher mass black holes, opening up
possibilities for light seed formation mechanisms, such as remnants of
population III stars, of being the progenitors of today's supermassive
black holes \citep{mad01,vol10}.

\subsection{Mass Retention Rate}\label{sec:retrate}

The mass retention rate ($\epsilon_{\rm m}$) is a free parameter in
the torque-limited model that we have introduced in order to match the
observed \Mbulge~relation (Sections~\ref{sec:growth}
and~\ref{sec:newmod}).  Figure~\ref{fig:eff} shows how the inferred
\Mbulge~relation changes when using a wide range of retention values,
from a extreme situation in which 99.9\,\% of the inflow mass is lost
in outflows, to the upper limit $\epsilon_{\rm m} = 1$ in which all of
the infalling gas from larger scales is accreted by the black hole
\citep[which corresponds to the original normalization of][]{hop11}.
As we might expect, we find that $\epsilon_{\rm m}$ only affects the
normalization of the \Mbulge~relation, with little effects on the
slope except at very early times.

A somewhat similar torque-limited scenario has been proposed by
\citet{esc06} and tested with idealized sub-parsec resolution
simulations of a nuclear galactic disk \citep{esc07}.  They find that
a mass retention rate $\epsilon_{\rm m} \approx 15$\,\% is required in
order to match the \Mbulge~relation, similar to our result
$\epsilon_{\rm m} \approx 5$\,\% given that both sets of simulations
probe spatial and timescales that are substantially different.

\begin{figure*}
\begin{center}
\includegraphics[scale=1]{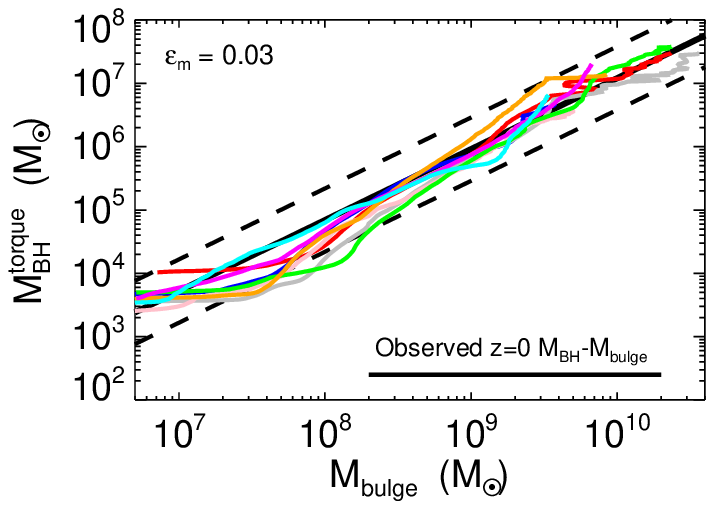}
\includegraphics[scale=1]{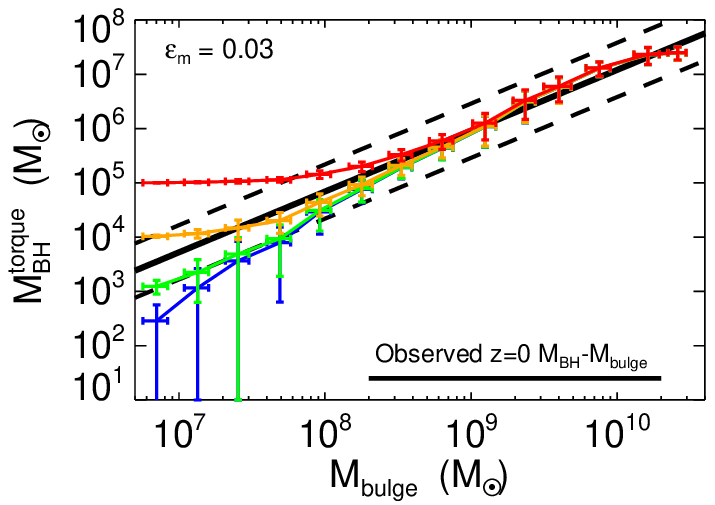}
\end{center}
\caption{Effects of stellar feedback on the \Mbulge~relation. Black holes 
grow according to the gravitational torque rate as in
Figure~\ref{fig:NumRel} but for simulations that do not include
galactic winds.  {\it Left}: initial black hole masses are taken to be
consistent with the \Mbulge~relation for each galaxy.  Each color
corresponds to a different galaxy as in Figure~\ref{fig:evol}.  {\it
Right}: average black hole mass within logarithmically spaced bins in bulge
mass, for all simulated galaxies, at all time steps, and for initial
black hole masses ranging from $10^{2}$\Msun~to $10^{5}$\Msun~(from blue to
red).  The mass retention rate $\epsilon_{\rm m} = 0.03$ has been used.
Black solid and dashed lines correspond to the observed
\Mbulge~relation \citep{har04} and a 0.5 dex scatter in black hole mass,
respectively.}
\label{fig:nw}
\end{figure*}

\begin{figure*}
\begin{center}
\includegraphics[scale=1]{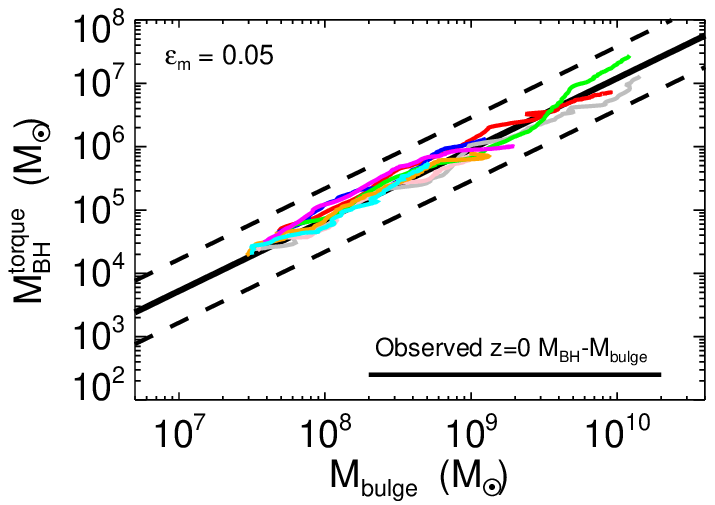}
\includegraphics[scale=1]{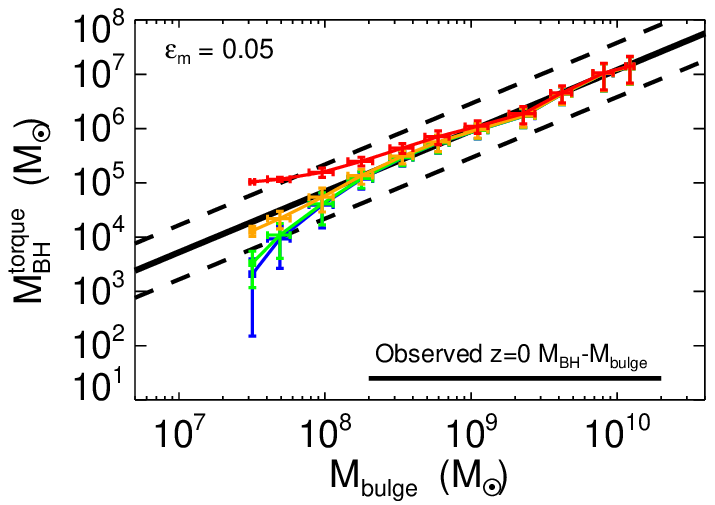}
\end{center}
\caption{Effects of numerical resolution on the \Mbulge~relation. Black holes 
grow according to the gravitational torque rate as in
Figure~\ref{fig:NumRel} but for simulations with a factor of two lower
spatial resolution and a factor of eight lower mass resolution.  Lines and
colors are as in Figure~\ref{fig:nw}.  The mass retention rate
$\epsilon_{\rm m} = 0.05$ has been used.}
\label{fig:lowres}
\end{figure*}

\subsection{Stellar Feedback and Resolution Effects}

In a separate work using the same set of simulations, we have shown
that strong stellar feedback in the form of galactic outflows can have
a significant impact on the star formation, morphological, and
kinematic properties of high-redshift galaxies \citep{ang12}.
Feedback from star formation could therefore affect black hole growth
\citep[e.g.,][]{cen12} and possibly the inferred black hole--galaxy
correlations.

Figure~\ref{fig:nw} shows the \Mbulge~relation obtained for the same
sample of galaxies but this time from simulations that do not include
models for galactic winds.  We find that the scaling relations are
still reproduced by the torque-limited model and perhaps even with
smaller scatter.  The convergence toward the \Mbulge~relation for
different initial black hole masses is retained (Section~\ref{sec:ini}) and
similar results are obtained for the \Msig~relation.  Predictions of
the gravitational torque model are therefore robust to changes in
stellar feedback.  Even if galaxies from simulations with no winds
become significantly more massive and less gas rich owing to higher
SFRs \citep{ang12}, black holes still grow according to the scaling relations
in the redshift range $z = 8 \rightarrow 2$.  This suggests that
galaxies regulate themselves via an equilibrium between inflows and
outflows and black holes grow according to the properties of their host
galaxies, at a rate given by the gas supply from galactic scales.

We have also investigated the effects of numerical resolution on the
scaling relations.  Figure~\ref{fig:lowres} shows the \Mbulge~relation
obtained for the same sample of galaxies from simulations carried out
with a factor of two lower spatial resolution and a factor of eight lower
mass resolution (including galactic outflows).  Since we require
galaxies to contain at least 100 gas particles and 100 star particles
within $R_{0} = 1$\,kpc in order to resolve their central morphology,
these lower resolution simulations can follow the evolution of black holes
only after their parent galaxies are a factor of $\sim 8$ more massive
compared to the high resolution simulations.  Despite the reduced
resolution, the \Mbulge~relation arising from the gravitational torque
model shows good numerical convergence.

We note that the values of $\epsilon_{\rm m}$ obtained for simulations
with no galactic outflows and the low-resolution simulations are
slightly different from our fiducial runs but they are all consistent
with a typical mass retention rate of a few percent of the gas supply.

\subsection{Black Hole Mergers}

It has been shown recently that black hole--galaxy correlations may be a
natural consequence of hierarchical structure formation, so that an
initially uncorrelated distribution of black hole and stellar masses tends to
converge toward the scaling relations through successive mergings of
black holes and galaxies \citep{pen07,jah11}.  In this scenario, AGN feedback
and self-regulated growth are not required, but it is assumed that
galaxy mergers always result in the merging of their central black holes.
Here, we have shown how torque-limited growth yields black hole--galaxy
correlations regardless of the initial masses of black holes and galaxies, and
also without the need for AGN feedback and self-regulated growth.  We
have, however, left off the discussion the role of black hole mergers in the
overall black hole growth.  Here, we assume that most of the black hole mass comes
from accretion \citep{sol82} and we simply neglect the mass
contribution from mergers.  Given that only a fraction of mass in our
galaxies comes from major mergers \citep{mura02,ker05}, we do not
expect a significant shift on the inferred black hole--galaxy correlations due
to black hole mergers.

\subsection{Evolution of the Scaling Relations}
 
Recent observations of active galaxies seem to indicate an evolution
of the \Mbulge~(and perhaps \Msig) relation with redshift, with black holes
being more massive for a given galaxy mass (or velocity dispersion) at
higher redshifts compared to the local relations
\citep{she08,woo08,dec10,gree10,mer10}.  There are also findings of no
significant evolution \citep{jah09} and potentially even undermassive
black holes in $z \sim 2$ infrared-selected galaxies \citep{shap09}.  In
either case, evaluations of the redshift evolution of the black hole scaling
relations may be biased by selection effects \citep{lau07}.

In the context of torque-limited growth, we find no evidence for
evolution of the scaling relations at least down to $z = 2$, since black holes
and galaxies seem to converge toward the scaling relations regardless
of their initial masses, provided $\alpha_{\rm T}$ and $\epsilon_{\rm
m}$ do not evolve with redshift and are well represented by a constant
factor.  Therefore, significant deviations at early times could in
principle reflect the ``initial conditions" for co-evolution of black holes and
galaxies (Figure~\ref{fig:ini}).

We note that $\epsilon_{\rm m}$ has been fixed here to match the 
\Mbulge~relation of \citet{har04} for the total stellar mass within 
the effective radius rather than the stellar bulge mass.  Given that
our galaxies contain a significant stellar disk component at $z = 2$
\citep{ang12}, it is plausible that there is significant evolution of
the \Mbulge~relation with respect to bulge mass but not with respect
to total stellar mass, as suggested by \citet{jah09}.  If most of the
mass of bulges and elliptical galaxies today comes from redistribution
of the stellar bulge+disk of $z = 2$ galaxies (through mergers or
other processes) no further black hole accretion would be required to match
the local scaling relations.  We speculate that the formation of
compact bulges would indeed decrease the gravitational torque rate
(since it is proportional to the disk mass), truncating significant black hole
growth at late times.


\section{Summary and Conclusions}\label{sec:end}

We have used cosmological zoom simulations of galaxy formation down to
$z = 2$, together with analytic models of black hole accretion, to investigate
the growth of massive black holes at the centers of galaxies without making
any prior assumptions about the effects of AGN feedback.  To this end,
we have compared predictions from the spherical Bondi model and an
accretion model driven by gravitational torques \citep{hop11}.

We find that the Bondi model presents significant challenges due to
the strong dependence of the accretion rates on black hole mass. Black hole growth is
significantly suppressed for low-mass black holes and becomes extremely rapid
for sufficiently massive black holes.  This has two main implications for
simulations of galaxy formation: (1) the initial black hole mass
(together with any boost factor or normalization constant) has to be
sufficiently massive to allow for early black hole growth and (2)
feedback energy and/or momentum needs to be injected into the
surrounding gas in order to regulate black hole growth at later times.
Because these implications follow from the $\propto M_{\rm BH}^2$
dependence, they should apply to any (reasonable) modification of the
original Bondi--Hoyle--Littleton parameterization.  Remarkably,
suitable choices of model parameters for different implementations
have been successful in reproducing a broad number of observations,
including the black hole--galaxy correlations
\citep[e.g.,][]{dimat05,rob06,hop07,boo09,joh09}.  

In the gravitational torque model, gas inflows are driven by global
gravitational instabilities in the disk and, therefore, do not depend
strongly on the black hole mass \citep{hop11}.  The resulting accretion rates
imply that (1) early black hole growth could in principle proceed at
super-critical rates, since the inflow rates can be well above the
Eddington limit even for small initial black holes, and (2) energy
(and/or momentum) from the accretion process does not need to couple
to galaxy-scale gas in order to regulate black hole growth.  
Consistency with the black hole--galaxy correlations simply requires the
assumption that only a small (constant) fraction of the mass inflow is
retained in the accretion flow, with the rest lost to winds and
outflows.  We have shown that this result is insensitive to variations
in the initial black hole mass, stellar feedback, or other implementation
details, and the required mass retention rate ($\epsilon_{\rm m}
\approx 5$\,\%) is roughly consistent with observational
\citep[e.g.,][]{king13} and theoretical expectations \citep[e.g.,][and
references therein]{yua12}.  However, the exact value of the
normalization factor $\epsilon_{\rm m}$ that we infer is subject to
uncertainties such as variations in the nuclear star formation law or
the exact form and normalization of the \Mbulge~relation \citep[see, e.g.,][]{gra12}.

Mass outflows are invoked in the torque-limited model in order to
control what fraction ($\epsilon_{\rm m}$) of the accretion disk
feeding rate is finally accreted by the central black hole (a linear effect),
but not to regulate the amount of gas feeding the accretion disk from
galactic scales.  In this scenario, there is no coupling between inflows and
outflows and, therefore, black hole feedback cannot shut down accretion.  ``Feedback self-regulation" does not occur in the strict sense of the term, since the energy output by accretion onto the black hole has no direct impact on its own fuel supply.  
This greatly differs from the non-linear feedback loop required by the Bondi model.  
Large-scale AGN feedback may have a significant impact on the host galaxy but it is not required for regulating black hole growth, which may instead be limited by the efficiency of gravitational torques in removing angular momentum from the gas together with competition with star formation.  

In addition, torque-limited growth yields a less direct correspondence
between major merger events and enhance AGN activity, as suggested by
recent observations of $z \sim 2$ active galaxies
\citep{koc12,mull12b,rosario12,scha12}.  Instead, black hole and galaxy growth are
governed by cosmological infall and transport of angular momentum in
the galactic disk, giving rise to a time-averaged connection between
AGN activity and SFR on cosmological timescales.  Our findings are
consistent with recent observational evidence for a constant black hole
accretion to star formation ratio, similar to the ratio of black hole mass
to stellar mass as inferred from the local \Mbulge~relation, and
independent of redshift since $z \sim 2$ \citep{raff11,mull12a}.

The agreement of feedback-regulated accretion models in hydrodynamic
simulations to the observed black hole--galaxy correlations has been
interpreted as (indirect) observational evidence of both AGN feedback
acting at galactic scales and the self-regulated growth of black holes.  We
note, however, that the black hole--galaxy correlations are indeed a primary
constraint for the strength of such feedback models \citep[see][for a
self-regulated model that accounts for angular momentum
transport]{deb11}.  Given that AGN feedback can have profound
consequences on the evolution of the host galaxy
\citep[e.g.,][]{dimat05} and that the scaling relations are not
necessarily a consequence of feedback, it would be desirable to
constrain AGN feedback models by other means, ideally through direct
observations of mass outflow rates \citep[e.g.,][]{feru10} and/or more
physically motivated models \citep[e.g.,][]{cio07,nov11,nov12}.  While
the Bondi model, appropriately tuned, does provide a good match to the
black hole--galaxy correlations, our work demonstrates that the torque-limited
model provides a viable alternative that does not require
self-regulation.

Our findings should apply so long as the assumptions implied by the
gravitational torque model are met.  In particular, it is assumed that
(1) there is a significant stellar component that drives the gas
into shocks that dissipate energy and angular momentum and (2)
the amplitudes of non-axisymmetric modes are large enough to produce
such shocks even at $\sim 10$\,pc scales \citep{hop11}.  These scales
are, of course, not resolved in our simulations; galaxy centers become
quickly stellar dominated but the formation time of the first star
particles is poorly constrained.  Other processes such as scattering
of dense gas clouds or the transport of angular momentum by supernova
and/or gravitational instability-driven turbulence may be required in
the very high gas-reach domain \citep[e.g.,][]{esc06,bour11,hobb11,gaspari13}.
Finally, full cosmological simulations with adequate resolution down
to $z = 0$ are required in order to evaluate the implications of
torque-limited growth on a statistically significant sample of
galaxies, making connection with the wealth of data available in the
local and low-redshift universe over a range of galaxy masses spanning
several orders of magnitude.

\acknowledgments

We thank K. Finlator, P. Hopkins, B. Oppenheimer, D. Psaltis, and
B. Robertson for stimulating conversations, and the anonymous referee for constructive comments that helped improve the paper. 
D.A.-A. thanks B. Robertson for very useful assistance with data visualization.  
F.\"O. gratefully acknowledges support from the Radcliffe Institute for Advanced Study at Harvard University.
The simulations were run 
on the University of Arizona's 512-processor SGI Altix system and the
TACC Sun Constellation Cluster (Ranger) at The University of Texas,
Austin. This work used the Extreme Science and Engineering Discovery
Environment (XSEDE), which is supported by National Science Foundation
grant number OCI-1053575.  This work was supported by the National
Science Foundation under grant numbers AST-0907998 and AST-1108753.
Computing resources were obtained through grant number DMS-0619881
from the National Science Foundation.

\end{document}